\documentclass[reviewcopy]{elsarticle}

\usepackage[reviewcopy]{adndt}
\usepackage{longtable}

\usepackage{amsmath}

\usepackage{amssymb}

\biboptions{square,sort&compress}
\bibpunct[]{[}{]}{,}{n}{}{;}
\begin{document}

\begin{frontmatter}

\journal{Atomic Data and Nuclear Data Tables}


\title{Radiative rates for  E1, E2, M1, and M2 transitions in Ne-like Hf LXIII, Ta LXIV and Re LXVI}

  \author[One]{Kanti M. Aggarwal\fnref{}\corref{cor1}}
  \ead{K.Aggarwal@qub.ac.uk}



  \cortext[cor1]{Corresponding author.}

  \address[One]{Astrophysics Research Centre, School of Mathematics and Physics, Queen's University Belfast,\\Belfast BT7 1NN,
Northern Ireland, UK}


\date{16.12.2002} 

\begin{abstract}  

Calculations for energy levels, radiative rates and lifetimes have been performed for three Ne-like ions, namely Hf~LXIII, Ta LXIV and Re LXVI, for which the general-purpose relativistic atomic structure package (GRASP) has been adopted. Results are presented among the lowest 121 levels of these ions, which belong to the 
2s$^2$2p$^6$,  2s$^2$2p$^5$3$\ell$, 2s2p$^6$3$\ell$, 2s$^2$2p$^5$4$\ell$, 2s2p$^6$4$\ell$, and 2s$^2$2p$^5$5$\ell$ configurations, but CI (configuration interaction) has been considered among a much larger number of levels. No measurements for energy levels are available but  comparisons have been made with the  earlier  similar  theoretical results.  Additionally,  calculations have also been performed with the flexible atomic code (FAC) in order to assess the effect of (much) larger CI on energy levels.   \\ \\

{\em Received}: 5 March 2018, {\em Accepted}: 3 April 2018

\vspace{0.5 cm}
\end{abstract}

\end{frontmatter}




\newpage

\tableofcontents
\listofDtables
\listofDfigures
\vskip5pc


\section{Introduction}

Emission lines of several Ne-like ions (particularly of Fe~XVII and Ni~XIX) among lower Z elements (up to 36) have been prominently observed in a variety of astrophysical plasmas and have been useful for diagnostics \cite{fe17a, fe17b, ni19a, ni19b}, for which atomic data for  several parameters, such as  energy levels, radiative rates and collision strengths, are required. Therefore, a lot of attention has been paid to these ions -- see for example J\"{o}nsson et al. \cite{jon} and references therein. However, there is comparatively a paucity of similar data for ions with higher Z, although W~LXV has attracted maximum attention due to its importance as a wall material in the developing ITER project -- see \cite{w65} and references therein. Therefore, recently Singh et al. {\cite{sam} have reported energy levels, radiative rates (A-values) and lifetimes ($\tau$) for four Ne-like ions, namely Hf~LXIII, Ta~LXIV, W~LXV,  and Re~LXVI. Since we have already reported these data for W~LXV \cite{w65}, we focus our attention on the remaining three ions. 

Earlier work on a wide range of Ne-like ions with Z $\le$ 92 was undertaken by Zhang et al. \cite{zs1, zs2} who considered 89 levels of the 2s$^2$2p$^6$, 2s$^2$2p$^5$$3\ell$, 2s2p$^6$$3\ell$,  2s$^2$2p$^5$4$\ell$,  and 2s2p$^6$$4\ell$ configurations. However, they calculated limited results because their focus was on the calculations of collision strengths for resonance transitions, i.e. those from the ground to higher excited levels. Additionally, for brevity energy levels were reported for only some ions and the ones of present interest were excluded. Similarly, Quinet et al. \cite{pq} calculated energy levels and A-values for several Ne-like ions, up to Z = 92, but not for the ions of present interest, although they did report oscillator strengths (f-values) for 5 E2 (electric quadrupole), 6 M1 (magnetic dipole) and 6 M2 (magnetic quadrupole) transitions of Hf~LXIII. For any ion the most dominant transitions of interest are E1 (electric dipole) because of their (much) larger magnitudes. Therefore, practically the only results available in the literature are those of Singh et al. \cite{sam}, and unfortunately no measurements have yet been made for the energy levels of current ions of interest, except for two levels of Hf~LXIII by  Beiersdorfer \cite{pb} through the electron beam ion trap machine.

For the calculations, Singh et al. \cite{sam} adopted two independent atomic structure codes, namely  the general-purpose relativistic atomic structure package (GRASP) and  the flexible atomic code (FAC) of Gu \cite{fac}. Both of these codes are freely available on the websites {\tt http://amdpp.phys.strath.ac.uk/UK\_APAP/codes.html} and {\tt https://www-amdis.iaea.org/FAC/}, respectively. In both calculations they included CI (configuration interaction) among 64 configurations, namely 2s$^2$2p$^6$, 2s$^2$2p$^5$$3\ell$, 2s2p$^6$3$\ell$,  2s$^2$2p$^5$4$\ell$,  2s2p$^6$$4\ell$,  2s$^2$2p$^5$5$\ell$, 2s2p$^6$$5\ell$, 2s$^2$2p$^5$6s/p/d, 2s$^2$2p$^5$7s/p/d, (2s$^2$2p$^4$) 3s3p, 3s3d, 3p3d, 3s$^2$, 3p$^2$, 3d$^2$, 3s4$\ell$, 3s5$\ell$, 3p4$\ell$, 3p5$\ell$, 3p5$\ell$, 3d4$\ell$, and 3d5$\ell$, which generate 3948 levels (or configuration state functions, CSF) in total. Energies for the lowest 209 levels were reported, which belong to the 2s$^2$2p$^6$, 2s$^2$2p$^5$$n\ell$ ($n \le$ 7, $\ell \le$ g) and 2s2p$^6$$n\ell$  ($n \le$ 5) configurations. Based on these two  calculations, they have `assessed' their energy levels to be accurate to $\sim$0.5~Ryd. However,  we notice that for some of the levels (particularly the higher ones)  the two calculations differ by up to  $\sim$2~Ryd,  see for example the (2p$^5$) 7p and 7d levels of Ta~LXIV, W~LXV and Re~LXVI in their tables~2--4, or present Table~B. This  is clearly contrary to their conclusion. Since such a large difference in energy levels with these two different codes (i.e. GRASP and FAC) is normally not  found,  we have performed our own calculations with the same configurations, as adopted by them. Unfortunately, we note that some of their results with FAC cannot be reproduced, and hence the large discrepancies.  In addition, 
there are other reasons to perform yet another calculation, as discussed below.

The  listed 209 levels of Singh et al. \cite{sam}  are not the lowest in energy,  because some of the neglected levels from other configurations, such as 2p$^5$6f/g/h and 2p$^5$7f/g/h/i,  intermix with these. Similarly, in spite of including a reasonably  large CI the effect of additional configurations need to be tested, although the ions considered are comparatively heavy. Finally, they have listed A-values for only resonance transitions, whereas data for {\em all}  among the listed levels are (highly) desired for the modelling of plasmas. Similarly, they have not reported A-values for E2, M1 and M2 transitions of these ions, and these data are important for considering a complete model as well as for the determination of $\tau$. Therefore, there is scope for improvement, confirmation and extension of their calculations, and this is necessary for having confidence in the data as suggested by \cite{rev1, rev2}.  

\section{Energy levels}

For the calculations we adopt the same (GRASP0) version as by Singh et al. \cite{sam}, which was originally developed by Grant et al. \cite{grasp0} but has been extensively  modified and updated by P.H.~Norrington,  i.e. one of the authors.  The option of  `extended average level' (EAL) is used for the optimisation of the orbitals  in which a weighted (proportional to 2$j$+1) trace of the Hamiltonian matrix is minimised. This option provides comparable results as with other choices, such as of `average level', and has been tested earlier on a wide range of ions.  Furthermore, as for calculations on other ions, the contributions of higher relativistic operators, namely Breit and quantum electro-dynamic effects (QED), are also included. Inclusion of these  improves the accuracy  of calculated energies. The maximum effect is on the energy of the ground levels, up to 49~Ryd, but is below 5~Ryd on higher excited ones. 

In our calculations with  GRASP, we have included CI among the same 64 configurations as by Singh et al. \cite{sam}, and listed in Section~1. Substantial inclusion of additional  CI is not possible in the version of the code adopted here, although it is feasible through other versions, such as GRASP2K \cite{grasp2k}, as demonstrated  by J\"{o}nsson et al. \cite{jon}. Nevertheless, we assess the effect of additional CI in our calculations with FAC, which generally provides comparable results as stated already, demonstrated in many of our earlier papers and further discussed below.

With FAC we have performed a series of calculations with increasing CI, but focus on only three, i.e.  (i) FAC1, which includes the same 3948 levels as by Singh et al. \cite{sam} and with GRASP,  (ii) FAC2, which includes 17~729 levels arising from all possible combinations of  2*8, (2*7) 3*1, 4*1, 5*1, 6*1, 7*1, (2*6) 3*2, 3*1 4*1, 3*1 5*1, 3*1 6*1, and 3*1 7*1,  and finally (iii) FAC3, which includes a total of 93~437 levels, the additional ones arising from  (2*6) 4*1 5*1, 4*1 6*1, 4*1 7*1, 5*1 6*1, 5*1 7*1, 6*1 7*1, and 2*5 3*3. The energy spans of the levels in these calculations are (almost) comparable. For example, for Re~LXVI the energy ranges of the three FAC calculations are about 2000, 2100 and 2500~Ryd, respectively. 

In Tables~1--3 we list energies for the levels of Hf~LXIII, Ta~LXIV and Re~LXVI, respectively. The listings in these tables are restricted to the lowest 121 levels of Hf~LXIII and Ta~LXIV and 117 of Re~LXVI. This is because beyond these levels from other configurations, such as 2p$^5$6$\ell$, intermix, but results for a larger range of levels can be obtained on request from the author. The listed energies are from our and earlier \cite{sam} calculations with GRASP, which are based on 3948 CSFs of 64 configurations. Also included in these tables are our final (FAC3) results with FAC which will help in assessing the contribution of larger CI. Before we discuss these energies in detail we will like to mention that the LSJ$^{\pi}$ designations are (a bit) ambiguous for about a dozen levels in each ion, because eigenvector from a particular level/configuration dominates for more than one. This is a common problem in all atomic  structure calculations, and the mixing coefficients for each level have already been provided by Singh et al. \cite{sam} in their tables~1, 2 and 4. Finally, the level orderings from GRASP and FAC are (mostly) similar with only a few exceptions, and we have tried our best  to match correspondence between the two calculations, although it is not straightforward  because of different levels of CI included and the designations (nomenclatures) provided by the two codes. 

In Table~A we compare our energies from three FAC calculations for the lowest 50 levels of Hf~LXIII, which will give an idea about the differences (or similarities) and hence an assessment of accuracy. The FAC1 and FAC2 energies for a few levels (such as 20--37) differ by a maximum of 0.1~Ryd. Considering that FAC2 calculations include larger CI,  by more than a factor of four, such discrepancies for a few levels are insignificant. Similarly, differences between the FAC2 and FAC3 energies are below 0.05~Ryd for a few levels, and much less for most, in spite of the latter having CI larger by more than a factor of five. This means that for the lower levels of Hf~LXIII (and subsequently of Ta~LXIV and Re~LXVI) inclusion of CI more than what has already been included in FAC1 (or GRASP) calculations is not beneficial, as the energies have (almost) converged for most levels.

Considering all levels listed in Tables~1--3, the maximum discrepancies between our GRASP and FAC calculations are up to  0.7~Ryd, but for only four (93-94 and 100-101),  and much less for most. This is highly satisfactory and we may therefore confidently state that both calculations (codes) provide comparable energies, as expected, and the listed results are accurate to better than 0.7~Ryd. In contrast, the energies listed by Singh et al. \cite{sam} with these codes, and with the same CI, differed by up to $\sim$2~Ryd for several levels, particularly the higher ones, as stated in Section~1 and seen in their tables~1--4. We will like to clarify here that we observe {\em no} such discrepancies among levels higher than 121, although results are not being listed here. However, in Table~B we demonstrate this for all four Ne-like ions. Included in this table are our and earlier \cite{sam} results with both GRASP and FAC, but only for four levels each of the (2p$^5$) 7p and 7d configurations, for which the discrepancies are the maximum.  For all the levels listed here energies among all our calculations agree within 0.5~Ryd, but those of Singh et al. \cite{sam} from FAC clearly stand out, and appear to be in error. Among the lowest 121 levels their corresponding differences are up to 1~Ryd (see levels 100-101 in their table~1), comparable to our calculations. However, it is interesting to note that our and their energies with GRASP differ by up to 0.35~Ryd for some levels and for all three ions -- see for example levels 22--27, 32--37 and 108--117 in Tables~1--3. For most such levels the energies obtained by Singh et al. are larger, and this is in spite of using the same CI and the code. We note here in passing that their calculations with both the GRASP and FAC codes had similar discrepancies  in the past for Br-like ions \cite{brlike} with 38 $\le$ Z $\le$ 42,  F-like W~LXVI \cite{w66a, w66b}, and very recently a few Ne-like \cite{nelike1}.

As stated in Section~1, measurements of energy are available \cite{pb} for two lines of Hf~LXIII, namely 3B (2s$^2$2p$^6$~$^1$S$_0$ -- 2s2p$^6$~$^3$P$^o_1$) and 3C (2s$^2$2p$^6$~$^1$S$_0$ -- 2s$^2$2p$^5$3d~$^1$P$^o_1$), whose upper levels correspond to 26 and 17, respectively, in Table~1. However, levels 17 (2s$^2$2p$^5$3d~$^1$P$^o_1$) and 29 (2s$^2$2p$^5$3d~$^3$D$^o_1$) are {\em mixed} with the coefficients 0.793~$^1$P$^o_1$$-$0.439~$^3$D$^o_1$ and 0.680~$^3$D$^o_1$+0.582~$^1$P$^o_1$, respectively, which are similar to those listed by \cite{sam}, and are clearly identifiable on the basis of the dominance of their coefficients. Therefore, the measurement of the 3C line corresponds to the J=1 level $^3$D$^o_1$. Anyway, the measured energies for these two levels are 710.3787$\pm$0.0514 and 714.8364$\pm$0.0514~Ryd, respectively, which compare favourably with our GRASP (710.5700 and 714.7552) and FAC results (710.5004 and 714.6761), listed in Table~1.

 \section{Radiative rates}\label{sec.eqs} 

Our calculated results  with the GRASP code  for transition energies (wavelengths, $\lambda_{ji}$ in ${\rm \AA}$), radiative rates (A-values, in s$^{-1}$), oscillator strengths (f-values, dimensionless), and line strengths (S-values, in atomic units, 1 a.u. = 6.460$\times$10$^{-36}$ cm$^2$ esu$^2$)  for  E1 transitions are listed in  Tables 4--6 for Hf~LXIII, Ta~LXIV and Re~LXVI, respectively.  These results are listed for {\em all} transitions among the  levels given in Tables~1--3. However,  for brevity only transitions from the lowest 3 to higher excited levels are listed in Tables~4--6, but  full tables in the ASCII format are available online in the electronic version. Also included in these tables are A-values for E2, M1 and M2 transitions, and  the corresponding data for f- or S-values can be  obtained using Eqs. (1-5) given in \cite{fe17b}.   Finally, calculations have been made  in both the  velocity (Coulomb) and length (Babushkin) gauges, but those from the latter are listed alone. Nevertheless, their  ratio (R) is also included in these tables, but only for E1 transitions which are the most dominant in magnitude and hence very important, for both plasma modelling and further calculations of lifetimes ($\tau$) -- to be discussed below.   For a majority of  strong transitions (with f $\ge$ 0.01) R is close(r) to unity, and this gives an indication about the accuracy of the results. However,  deviations from unity are sometimes  large(r) for a few  weak ones, because of the additive or cancellation effects of the contributing vectors -- see \cite{rev2} for further discussion and examples.

The only results available in the literature with which to make comparisons are those of Singh et al. \cite{sam} for some E1 transitions, but for all ions, and of Quinet et al. \cite{pq} for a few E2, M1 and M2 of Hf~LXIII alone. In Table~C we compare our f-values with those of \cite{sam} for the common E1 transitions of Hf~LXIII, for which there are no appreciable discrepancies. This result is expected because both calculations adopt the same code and include the same CI, and small differences (if any) in energy levels do not substantially affect the subsequent calculations of f- or A-values. Similar comparisons are shown in Tables~D, E and F for the E2, M1 and M2 transitions, respectively, with the earlier calculations of Quinet et al. Again,  there are no appreciable differences between the two calculations, in spite of the fact that  the inclusion of CI is very (very) different between the two calculations. However, considering that there are 2077~E1, 2653~E2, 2026~M1, and 2701~M2 possible transitions among the 121 levels listed in Table~1, this comparison is very limited. Nevertheless, we compare the lifetimes in the next section, which will give further idea about the accuracy of our radiative rates.

\section{Lifetimes}

The lifetime $\tau$ of a level $j$ is defined as 1.0/$\Sigma_{i}$A$_{ji}$, where the summation runs  over the A-values for all types of transitions, i.e. E1, E2, M1, and M2.  Although the E1 transitions are (generally) the most dominant, as already stated, contributions from others not only improve the accuracy but are very helpful for those levels for which the former do not connect, such as level 2, i.e. 2s$^2$2p$^5$3s~$^3$P$ ^o_{2}$.  Similarly, it is a measurable quantity and hence can help in assessing the accuracy of theoretical results, but to the best of our knowledge no experiments have yet been performed for transitions/levels of the ions under consideration, and the only theoretical results available for comparison are those of Singh et al. \cite{sam}, who have used the same code, methodology and CI. Therefore, in Table~G we compare our results with theirs for the lowest 50 levels of all three ions, i.e. Hf~LXIII, Ta~LXIV and Re~LXVI. This comparison should  give some idea about the  accuracy.  As expected, there are no notable discrepancies between the two sets of results, and the only  level which points to some disagreement  is number 4 (2s$^2$2p$^5$3p~$^3$P$  _{1}$) of Re~LXVI for which the earlier value is lower by a factor of 15. There are two transitions which  `effectively' contribute to this level, namely 1--4~M1 and 2--4~E1 with the A-values 3.54$\times$10$^{10}$   and 3.09$\times$10$^{10}$~s$^{-1}$, respectively. Unfortunately, only for the former the A-values can be compared (see Table~E) for which there is no discrepancy. Additionally, since this is the only level (among all three ions) for which the differences are (rather) large, the reported result of Singh et al. does not appear to be correct. We will also like to note here that similar discrepancies for a level or two have also been noted in their work on other ions -- see for example \cite{kmq, ba48} for F-like ions. Overall, for most levels there are no discrepancies between the two sets of results, and hence we may state that our calculated values of $\tau$ (and hence of the A-values too) are  accurate to  $\sim$20\%. Nevertheless, measurements for (at least) a few levels as well as other theoretical work/s will be helpful in further assessing the accuracy of our calculated results.

\section{Conclusions}

We have calculated energies for the levels of three Ne-like ions, namely Hf~LXIII, Ta~LXIV and Re~LXVI, for which the GRASP code has been used, and CI among 64 configurations with $n \le$ 7 has been considered. These configurations generate 3948 CSFs (levels) in total but for brevity results have been listed for the lowest 121 alone, which belong to  2s$^2$2p$^6$,  2s$^2$2p$^5$3$\ell$, 2s2p$^6$3$\ell$, 2s$^2$2p$^5$4$\ell$, 2s2p$^6$4$\ell$, and 2s$^2$2p$^5$5$\ell$. Although earlier similar results are available for comparison, calculations have also been performed with FAC, and with much larger CI including up to 93~437 CSFs. This is for two reasons, firstly to assess the significance of larger CI, and secondly to have confidence in the data, because no measurements are available for any of the ions under consideration. The effect of larger CI is found to be insignificant on these ions with high Z, and based on several calculations as well as comparisons with existing results, our energies are estimated to be accurate to better than 0.7~Ryd for all levels and ions.

Results have also been calculated for A-values, and for four types, i.e. E1, E2, M1, and M2. Only limited results (for a few transitions) are available in the existing literature with which comparisons could be made, but no appreciable discrepancies have been noted. We hope our complete set of listed results for a large number of transitions will be useful for plasma modelling. Additionally, with these calculated A-values, lifetimes have also been determined, although no measurements have so far been performed. However, our results are comparable with those already available, and hence are assessed   to be  accurate to $\sim$20\%.




\begin{appendix}

\def\thesection{} 

\section{Appendix A. Supplementary data}

Owing to space limitations, only parts of Tables 4--6  are presented here, but full tables are being made available as supplemental material in conjunction with the electronic publication of this work. Supplementary data associated with this article can be found, in the online version, at doi:nn.nnnn/j.adt.2018.nn.nnn.

\end{appendix}



\newpage
\renewcommand{\baselinestretch}{1.0}
\footnotesize
\begin{longtable}{@{\extracolsep\fill}rllrrrrrrrrr@{}}
\caption{Comparison of energies (in Ryd) with the FAC code for the lowest 50 levels of  Hf LXIII.}
Index  &     Configuration 	& Level 	     &      FAC1    &  FAC2    &    FAC3    \\  \\
\hline \\
\endfirsthead\\
\caption[]{(continued)}
Index  &     Configuration 	& Level 	     &     FAC1    &  FAC2     &      FAC3 \\  \\
\hline \\

\endhead 
    1   &   2s$^2$2p$^6$	   &      $^1$S$  _{0}$   &    0.00000    &    0.0000	&    0.0000   \\
    2   &   2s$^2$2p$^5$3s	   &      $^3$P$^o_{2}$   &  575.85651    &  575.8300	&  575.7994   \\   
    3   &   2s$^2$2p$^5$3s	   &      $^1$P$^o_{1}$   &  576.46570    &  576.4261	&  576.3963   \\   
    4   &   2s$^2$2p$^5$3p	   &      $^3$P$  _{1}$   &  586.37079    &  586.3268	&  586.2974   \\ 
    5   &   2s$^2$2p$^5$3p	   &      $^3$D$  _{2}$   &  586.50519    &  586.4614	&  586.4324   \\ 
    6   &   2s$^2$2p$^5$3p	   &      $^1$P$  _{1}$   &  610.85870    &  610.8110	&  610.7811   \\
    7   &   2s$^2$2p$^5$3p	   &      $^3$D$  _{3}$   &  610.88629    &  610.8351	&  610.8049   \\
    8   &   2s$^2$2p$^5$3p	   &      $^3$P$  _{2}$   &  611.77429    &  611.7274	&  611.6983   \\
    9   &   2s$^2$2p$^5$3p	   &      $^1$S$  _{0}$   &  615.82037    &  615.7783	&  615.7249   \\
   10   &   2s$^2$2p$^5$3d	   &      $^3$P$^o_{0}$   &  622.10046    &  622.0540	&  622.0213   \\
   11   &   2s$^2$2p$^5$3d	   &      $^3$P$^o_{1}$   &  622.80560    &  622.7582	&  622.7255   \\
   12   &   2s$^2$2p$^5$3d	   &      $^3$F$^o_{3}$   &  622.89215    &  622.8472	&  622.8141   \\
   13   &   2s$^2$2p$^5$3d	   &      $^3$D$^o_{2}$   &  623.45392    &  623.4058	&  623.3733   \\
   14   &   2s$^2$2p$^5$3d	   &      $^3$F$^o_{4}$   &  628.25214    &  628.2138	&  628.1805   \\
   15   &   2s$^2$2p$^5$3d	   &      $^1$D$^o_{2}$   &  628.75073    &  628.6980	&  628.6664   \\
   16   &   2s$^2$2p$^5$3d	   &      $^3$D$^o_{3}$   &  629.46375    &  629.4122	&  629.3807   \\ 
   17   &   2s$^2$2p$^5$3d	   &      $^1$P$^o_{1}$   &  631.53229    &  631.4867	&  631.4429   \\ 
   18   &   2s$^2$2p$^5$3s	   &      $^3$P$^o_{0}$   &  665.65515    &  665.6173	&  665.5866   \\ 
   19   &   2s$^2$2p$^5$3s	   &      $^3$P$^o_{1}$   &  665.90765    &  665.8623	&  665.8317   \\ 
   20   &   2s$^2$2p$^5$3p	   &      $^3$D$  _{1}$   &  675.92041    &  675.8596	&  675.8298   \\ 
   21   &   2s$^2$2p$^5$3p	   &      $^3$P$  _{0}$   &  679.36182    &  679.3065	&  679.2625   \\ 
   22   &   2s2p$^6$3s  	   &      $^3$S$  _{1}$   &  698.86835    &  698.8243	&  698.7856   \\ 
   23   &   2s$^2$2p$^5$3p	   &      $^1$D$  _{2}$   &  700.97876    &  700.9153	&  700.8856   \\ 
   24   &   2s$^2$2p$^5$3p	   &      $^3$S$  _{1}$   &  702.08246    &  701.9921	&  701.9584   \\ 
   25   &   2s2p$^6$3s  	   &      $^1$S$  _{0}$   &  702.44794    &  702.3398	&  702.2870   \\ 
   26   &   2s2p$^6$3p  	   &      $^3$P$^o_{1}$   &  710.49969    &  710.4342	&  710.3934   \\ 
   27   &   2s2p$^6$3p  	   &      $^3$P$^o_{0}$   &  710.71338    &  710.6473	&  710.6068   \\ 
   28   &   2s$^2$2p$^5$3d	   &      $^3$F$^o_{2}$   &  712.66199    &  712.6045	&  712.5712   \\
   29   &   2s$^2$2p$^5$3d	   &      $^3$D$^o_{1}$   &  714.64178    &  714.5742	&  714.5359   \\
   30   &   2s$^2$2p$^5$3d	   &      $^3$P$^o_{2}$   &  718.48578    &  718.4211	&  718.3894   \\ 
   31   &   2s$^2$2p$^5$3d	   &      $^1$F$^o_{3}$   &  718.73468    &  718.6721	&  718.6398   \\ 
   32   &   2s2p$^6$3p  	   &      $^3$P$^o_{2}$   &  735.33466    &  735.2728	&  735.2315   \\ 
   33   &   2s2p$^6$3p  	   &      $^1$P$^o_{1}$   &  735.76788    &  735.6962	&  735.6557   \\ 
   34   &   2s2p$^6$3d  	   &      $^3$D$  _{1}$   &  746.98590    &  746.9171	&  746.8730   \\
   35   &   2s2p$^6$3d  	   &      $^3$D$  _{2}$   &  747.37042    &  747.2980	&  747.2539   \\
   36   &   2s2p$^6$3d  	   &      $^3$D$  _{3}$   &  752.61786    &  752.5529	&  752.5095   \\
   37   &   2s2p$^6$3d  	   &      $^1$D$  _{2}$   &  753.72119    &  753.6455	&  753.6021   \\
   38   &   2s$^2$2p$^5$4s	   &      $^3$P$^o_{2}$   &  804.57867    &  804.5428	&  804.5392   \\ 
   39   &   2s$^2$2p$^5$4s	   &      $^1$P$^o_{1}$   &  804.77893    &  804.7439	&  804.7404   \\ 
   40   &   2s$^2$2p$^5$4p	   &      $^3$P$  _{1}$   &  808.91718    &  808.8842	&  808.8815   \\
   41   &   2s$^2$2p$^5$4p	   &      $^3$D$  _{2}$   &  808.95660    &  808.9239	&  808.9212   \\
   42   &   2s$^2$2p$^5$4p	   &      $^3$D$  _{3}$   &  818.97052    &  818.9376	&  818.9345   \\
   43   &   2s$^2$2p$^5$4p	   &      $^1$P$  _{1}$   &  818.99286    &  818.9587	&  818.9556   \\
   44   &   2s$^2$2p$^5$4p	   &      $^3$P$  _{2}$   &  819.31006    &  819.2775	&  819.2744   \\
   45   &   2s$^2$2p$^5$4p	   &      $^1$S$  _{0}$   &  820.65308    &  820.6208	&  820.6194   \\ 
   46   &   2s$^2$2p$^5$4d	   &      $^3$P$^o_{0}$   &  823.28259    &  823.2486	&  823.2458   \\ 
   47   &   2s$^2$2p$^5$4d	   &      $^3$P$^o_{1}$   &  823.53528    &  823.5016	&  823.4988   \\
   48   &   2s$^2$2p$^5$4d	   &      $^3$F$^o_{3}$   &  823.55121    &  823.5176	&  823.5146   \\ 
   49   &   2s$^2$2p$^5$4d	   &      $^3$D$^o_{2}$   &  823.75281    &  823.7194	&  823.7163   \\ 
   50   &   2s$^2$2p$^5$4d	   &      $^3$F$^o_{4}$   &  825.89459    &  825.8603	&  825.8577   \\ 
\\  \hline            								                	 
\end{longtable}   								   					       
			      							   					       

														       
\begin{flushleft}													       
{\small
FAC1: Present calculations  with the FAC code  for 3948 levels \\
FAC2: Present calculations  with the FAC code  for 17~729 levels \\
FAC3: Present calculations  with the FAC code  for 93~437 levels \\   
}															       
\end{flushleft} 


\clearpage
\newpage 

\newpage
\renewcommand{\baselinestretch}{1.0}
\footnotesize
\begin{longtable}{@{\extracolsep\fill}rllrrrrrrrrr@{}}
\caption{Comparison of some energy levels  (in Ryd) of Ne-like ions.}
Level (2p$^5$) & 7p~$^3$D$_1$ & 7p~$^3$P$_0$ & 7p~$^3$S$_1$ & 7p~$^1$D$_2$ & 7d~$^3$F$^o_2$ & 7d~$^3$D$^o_1$ & 7d~$^3$P$^o_2$ & 7d~$^1$F$^o_3$    \\  \\
\hline \\
\endfirsthead\\
\caption[]{(continued)}
Level (2p$^5$) & 7p~$^3$D$_1$ & 7p~$^3$P$_0$ & 7p~$^3$S$_1$ & 7p~$^1$D$_2$ & 7d~$^3$F$^o_2$ & 7d~$^3$D$^o_1$ & 7d~$^3$P$^o_2$ & 7d~$^1$F$^o_3$ \\  \\
\hline \\

\endhead 
{\bf i.} Hf~LXIII \\ \\
 GRASP1a & 1083.665 & 1083.837 & 1085.502 & 1085.506 & 1086.274 & 1086.377 & 1086.750 & 1086.762 \\
 GRASP1b & 1083.651 & 1083.821 & 1085.489 & 1085.493 & 1086.261 & 1086.364 & 1086.737 & 1086.749 \\
 FAC1a   & 1083.972 & 1084.129 & 1084.972 & 1085.168 & 1085.835 & 1085.841 & 1086.250 & 1086.259 \\
 FAC1b   & 1083.707 & 1083.861 & 1085.554 & 1085.567 & 1086.350 & 1086.450 & 1086.831 & 1086.846 \\
 FAC2    & 1083.511 & 1083.668 & 1085.379 & 1085.372 & 1086.156 & 1086.264 & 1086.634 & 1086.651 \\
 FAC3    & 1083.466 & 1083.623 & 1085.334 & 1085.327 & 1086.110 & 1086.219 & 1086.588 & 1086.605 \\ \\
 \hline \\
 {\bf ii.} Ta~LXIV \\ \\
 GRASP1a & 1120.443 & 1120.620 & 1222.404 & 1122.407 & 1123.189 & 1123.293 & 1123.695 & 1123.707 \\
 GRASP1b & 1120.429 & 1120.603 & 1122.391 & 1122.393 & 1123.175 & 1123.279 & 1123.682 & 1123.693 \\
 FAC1a   & 1120.758 & 1120.918 & 1121.396 & 1121.590 & 1122.694 & 1122.707 & 1122.745 & 1122.745 \\
 FAC1b   & 1120.486 & 1120.644 & 1122.456 & 1122.469 & 1123.267 & 1123.368 & 1123.778 & 1123.793 \\
 FAC2    & 1120.290 & 1120.451 & 1122.276 & 1122.276 & 1123.073 & 1123.178 & 1123.582 & 1123.599 \\
 FAC3    & 1120.245 & 1120.406 & 1122.230 & 1122.230 & 1123.027 & 1123.133 & 1123.536 & 1123.554 \\ \\
 \hline \\
 {\bf iii.} W~LXV \\ \\
 GRASP1a & 1158.024 & 1158.205 & 1160.113 & 1160.116 & 1160.912 & 1161.017 & 1161.450 & 1161.462 \\
 GRASP1b & 1158.009 & 1158.187 & 1160.099 & 1160.102 & 1160.897 & 1161.003 & 1161.436 & 1161.448 \\
 FAC1a   & 1158.343 & 1158.509 & 1158.609 & 1158.803 & 1159.928 & 1159.941 & 1160.461 & 1160.464 \\
 FAC1b   & 1158.067 & 1158.228 & 1160.167 & 1160.180 & 1160.991 & 1161.094 & 1161.535 & 1161.551 \\
 FAC2    & 1157.869 & 1158.034 & 1159.984 & 1159.988 & 1160.798 & 1160.904 & 1161.339 & 1161.357 \\
 FAC3    & 1157.824 & 1157.989 & 1159.939 & 1159.943 & 1160.753 & 1160.859 & 1161.294 & 1161.312 \\ \\
 \hline \\
 {\bf iv.} Re~LXVI \\ \\
 GRASP1a & 1196.421 & 1196.606 & 1198.647 & 1198.650 & 1199.458 & 1199.565 & 1200.031 & 1200.042 \\
 GRASP1b & 1196.406 & 1196.588 & 1198.633 & 1198.635 & 1199.443 & 1199.551 & 1200.016 & 1200.028 \\
 FAC1a   & 1196.608 & 1196.768 & 1196.807 & 1196.934 & 1197.972 & 1197.985 & 1199.002 & 1199.008 \\
 FAC1b   & 1196.465 & 1196.629 & 1198.702 & 1198.715 & 1199.540 & 1199.644 & 1200.118 & 1200.134 \\
 FAC2    & 1196.290 & 1196.462 & 1198.518 & 1198.523 & 1199.347 & 1199.454 & 1199.922 & 1199.941 \\ 
 FAC3    & 1196.247 & 1196.421 & 1198.473 & 1198.479 & 1199.303 & 1199.409 & 1199.877 & 1199.896 \\ 
\\  \hline            								                	 
\end{longtable}   								   					       
			      							   					       

														       
\begin{flushleft}													       
{\small
GRASP1a: Earlier calculations  of Singh et al. \cite{sam} with the GRASP code for 3948 levels \\
GRASP1b: Present calculations  with the GRASP code for 3948 levels \\
FAC1a:  Earlier calculations  of Singh et al. \cite{sam} with the FAC code for 3948 levels \\
FAC1b: Present calculations  with the FAC code  for 3948 levels \\
FAC2: Present calculations  with the FAC code  for 17~729 levels \\
FAC3: Present calculations  with the FAC code  for 93~437 levels \\   
}															       
\end{flushleft} 


\clearpage
\newpage

\renewcommand{\baselinestretch}{1.0}
\footnotesize
\begin{longtable}{@{\extracolsep\fill}rllrrrrrrrrr@{}}
\caption{Comparison of oscillator strengths (f-values)  for some E1 transitions of Hf LXIII from the ground level.  $a{\pm}b \equiv$ $a\times$10$^{{\pm}b}$.}
J & GRASP$^a$ & GRASP$^b$  \\  \\
\hline \\
\endfirsthead\\
\caption[]{(continued)}
J & GRASP$^a$ & GRASP$^b$  \\  \\
\hline \\

\endhead 
  3    &  1.50-1  &   1.52-1   \\
 11    &  6.16-2  &   6.14-2   \\
 17    &  2.31-0  &   2.32-0   \\
 19    &  2.75-2  &   2.80-2   \\
 26    &  6.80-1  &   7.34-1   \\
 29    &  6.94-1  &   6.47-1   \\
 33    &  2.73-1  &   2.75-1   \\
 39    &  2.66-2  &   2.75-2   \\
 47    &  1.44-2  &   1.43-2   \\
 53    &  4.84-1  &   4.90-1   \\
 63    &  7.73-3  &   8.07-3   \\
 67    &  1.29-2  &   1.37-2   \\
 77    &  2.00-1  &   2.02-1   \\
 79    &  4.37-3  &   4.33-3   \\
 87    &  2.36-1  &   2.42-1   \\
 111   &  5.26-2  &   5.39-2   \\
 113   &  8.65-2  &   8.90-2   \\

\\  \hline            								                	 
\end{longtable}   								   					       
			      							   					       

														       
\begin{flushleft}													       
{\small					      		      
GRASP$^a$: Present calculations  with the GRASP code for 3948 levels \\
GRASP$^b$: Calculations of Singh et al. \cite{sam}  with the GRASP code for 3948 levels \\ 
}															       
\end{flushleft} 


\vspace*{1.0 cm} 

\renewcommand{\baselinestretch}{1.0}
\footnotesize
\begin{longtable}{@{\extracolsep\fill}rllrrrrrrrrr@{}}
\caption{Comparison of oscillator strengths (f-values)  for some E2 transitions of Hf LXIII from the ground level.  $a{\pm}b \equiv$ $a\times$10$^{{\pm}b}$.}
J & GRASP$^a$ & GRASP$^b$  \\  \\
\hline \\
\endfirsthead\\
\caption[]{(continued)}
J & GRASP$^a$ & GRASP$^b$  \\  \\
\hline \\

\endhead 
   5 &    1.40-3   &   1.4-3  \\  
   8 &    1.27-3   &   1.3-3  \\  
  23 &    1.17-3   &   1.2-3  \\  
  35 &    4.39-3   &   4.4-3  \\  
  37 &    1.09-2   &   1.1-2  \\  
\\  \hline            								                	 
\end{longtable}   								   					       
			      							   					       

														       
\begin{flushleft}													       
{\small
GRASP$^a$: Present calculations  with the GRASP code for 3948 levels \\ 
GRASP$^b$: Calculations of Quinet et al. \cite{pq}  with the GRASP code \\
}															       
\end{flushleft} 


\clearpage
\newpage

\renewcommand{\baselinestretch}{1.0}
\footnotesize
\begin{longtable}{@{\extracolsep\fill}rllrrrrrrrrr@{}}
\caption{Comparison of oscillator strengths (f-values)  for some M1 transitions of Hf LXIII from the ground level.  $a{\pm}b \equiv$ $a\times$10$^{{\pm}b}$.}
J & GRASP$^a$ & GRASP$^b$  \\  \\
\hline \\
\endfirsthead\\
\caption[]{(continued)}
J & GRASP$^a$ & GRASP$^b$  \\  \\
\hline \\

\endhead 
   4   &  2.45-5   &  2.5-5  \\      
   6   &  7.70-6   &  7.7-6  \\      
  20   &  7.62-7   &  8.0-7  \\      
  22   &  2.08-6   &  2.3-6  \\      
  24   &  1.33-5   &  1.3-5  \\      
  34   &  2.62-7   &  3.0-7  \\      
\\  \hline            								                	 
\end{longtable}   								   					       
			      							   					       

														       
\begin{flushleft}													       
{\small
GRASP$^a$: Present calculations  with the GRASP code for 3948 levels \\ 
GRASP$^b$: Calculations of Quinet et al. \cite{pq}  with the GRASP code \\
}															       
\end{flushleft} 


\vspace*{1.0 cm} 

\renewcommand{\baselinestretch}{1.0}
\footnotesize
\begin{longtable}{@{\extracolsep\fill}rllrrrrrrrrr@{}}
\caption{Comparison of oscillator strengths (f-values)  for some M2 transitions of Hf LXIII from the ground level.  $a{\pm}b \equiv$ $a\times$10$^{{\pm}b}$.}
J & GRASP$^a$ & GRASP$^b$  \\  \\
\hline \\
\endfirsthead\\
\caption[]{(continued)}
J & GRASP$^a$ & GRASP$^b$  \\  \\
\hline \\

\endhead 
    2  &  6.21-6    &  6.4-6  \\      
   13  &  3.52-7    &  4.0-7  \\      
   15  &  1.50-4    &  1.5-4  \\      
   28  &  2.81-6    &  2.9-6  \\      
   30  &  3.16-5    &  3.2-5  \\      
   32  &  1.89-5    &  1.9-5  \\      
\\  \hline            								                	 
\end{longtable}   								   					       
			      							   					       

														       
\begin{flushleft}													       
{\small
GRASP$^a$: Present calculations  with the GRASP code for 3948 levels \\ 
GRASP$^b$: Calculations of Quinet et al. \cite{pq}  with the GRASP code \\
}															       
\end{flushleft} 


\clearpage
\newpage

\renewcommand{\baselinestretch}{1.0}
\footnotesize
\begin{longtable}{@{\extracolsep\fill}rlrrrrrrrrrrr@{}}
\caption{Comarison of lifetimes ($\tau$, s) for the lowest 50 levels of  Hf LXIII, Ta LXIV and Re LXVI. $a{\pm}b \equiv$ $a\times$10$^{{\pm}b}$.}
Ion & & & \multicolumn{2}{c}{Hf LXIII} & \multicolumn{2}{c}{Ta LXIV} &\multicolumn{2}{c}{Re LXVI}   \\  \\
\hline \\
Index  &     Config.  & Level  & GRASP$^a$ & GRASP$^b$ &   GRASP$^a$ & GRASP$^b$ &   GRASP$^a$ & GRASP$^b$    \\
\\ \hline  \\  
\endfirsthead\\
\caption[]{(continued)}
Ion & \multicolumn{3}{c}{Hf LXIII} & \multicolumn{3}{c}{Ta LXIV} &\multicolumn{3}{c}{Re LXVI}   \\  \\
\hline \\
Index  &     Config.  & Level  & GRASP$^a$ & GRASP$^b$ &   GRASP$^a$ & GRASP$^b$ &   GRASP$^a$ & GRASP$^b$   \\
\\ \hline  \\  \\
\hline\\
\endhead 
    1   &   2s$^2$2p$^6$	   &      $^1$S$  _{0}$   &  ........     &  .......   &  ........     &  .......   &  ........     &	.......   \\
    2   &   2s$^2$2p$^5$3s	   &      $^3$P$^o_{2}$   &  3.023-10     &  2.99-10   &  2.656-10     &  2.63-10   &  2.060-10     &	2.03-10   \\
    3   &   2s$^2$2p$^5$3s	   &      $^1$P$^o_{1}$   &  7.476-15     &  7.41-15   &  6.988-15     &  6.93-15   &  6.115-15     &	6.07-15   \\
    4   &   2s$^2$2p$^5$3p	   &      $^3$P$  _{1}$   &  1.823-11     &  1.81-11   &  1.678-11     &  1.67-11   &  1.410-11     &	9.24-13   \\
    5   &   2s$^2$2p$^5$3p	   &      $^3$D$  _{2}$   &  1.238-12     &  1.23-12   &  1.136-12     &  1.13-12   &  9.599-13     &	9.56-13   \\
    6   &   2s$^2$2p$^5$3p	   &      $^1$P$  _{1}$   &  8.091-13     &  8.08-13   &  7.131-13     &  7.12-13   &  5.542-13     &	5.54-13   \\
    7   &   2s$^2$2p$^5$3p	   &      $^3$D$  _{3}$   &  7.685-13     &  7.68-13   &  6.788-13     &  6.79-13   &  5.299-13     &	5.30-13   \\
    8   &   2s$^2$2p$^5$3p	   &      $^3$P$  _{2}$   &  4.653-13     &  4.64-13   &  4.166-13     &  4.16-13   &  3.343-13     &	3.34-13   \\
    9   &   2s$^2$2p$^5$3p	   &      $^1$S$  _{0}$   &  5.139-13     &  5.11-13   &  4.593-13     &  4.57-13   &  3.668-13     &	3.65-13   \\
   10   &   2s$^2$2p$^5$3d	   &      $^3$P$^o_{0}$   &  1.032-12     &  1.04-12   &  9.172-13     &  9.21-13   &  7.246-13     &	7.28-13   \\
   11   &   2s$^2$2p$^5$3d	   &      $^3$P$^o_{1}$   &  1.538-14     &  1.54-14   &  1.359-14     &  1.36-14   &  1.074-14     &	1.08-14   \\
   12   &   2s$^2$2p$^5$3d	   &      $^3$F$^o_{3}$   &  9.772-13     &  9.82-13   &  8.707-13     &  8.75-13   &  6.908-13     &	6.94-13   \\
   13   &   2s$^2$2p$^5$3d	   &      $^3$D$^o_{2}$   &  9.292-13     &  9.34-13   &  8.288-13     &  8.33-13   &  6.589-13     &	6.62-13   \\
   14   &   2s$^2$2p$^5$3d	   &      $^3$F$^o_{4}$   &  7.266-12     &  7.31-12   &  6.805-12     &  6.84-12   &  5.961-12     &	5.99-12   \\
   15   &   2s$^2$2p$^5$3d	   &      $^1$D$^o_{2}$   &  4.178-12     &  4.19-12   &  3.831-12     &  3.84-12   &  3.217-12     &	3.22-12   \\
   16   &   2s$^2$2p$^5$3d	   &      $^3$D$^o_{3}$   &  6.615-12     &  6.65-12   &  6.205-12     &  6.24-12   &  5.452-12     &	5.48-12   \\
   17   &   2s$^2$2p$^5$3d	   &      $^1$P$^o_{1}$   &  4.052-16     &  4.04-16   &  3.818-16     &  3.80-16   &  3.400-16     &	3.39-16   \\
   18   &   2s$^2$2p$^5$3s	   &      $^3$P$^o_{0}$   &  4.484-11     &  4.53-11   &  3.743-11     &  3.78-11   &  2.624-11     &	2.65-11   \\
   19   &   2s$^2$2p$^5$3s	   &      $^3$P$^o_{1}$   &  3.059-14     &  3.01-14   &  2.816-14     &  2.77-14   &  2.409-14     &	2.37-14   \\
   20   &   2s$^2$2p$^5$3p	   &      $^3$D$  _{1}$   &  1.844-11     &  1.84-11   &  1.679-11     &  1.68-11   &  1.374-11     &	1.38-11   \\
   21   &   2s$^2$2p$^5$3p	   &      $^3$P$  _{0}$   &  3.717-12     &  3.53-12   &  3.435-12     &  3.27-12   &  2.916-12     &	2.77-12   \\
   22   &   2s2p$^6$3s  	   &      $^3$S$  _{1}$   &  1.076-13     &  1.19-13   &  7.255-14     &  7.79-14   &  4.344-14     &	4.45-14   \\
   23   &   2s$^2$2p$^5$3p	   &      $^1$D$  _{2}$   &  4.445-13     &  4.43-13   &  3.977-13     &  3.97-13   &  4.075-14     &	4.10-14   \\
   24   &   2s$^2$2p$^5$3p	   &      $^3$S$  _{1}$   &  9.136-14     &  8.52-14   &  5.224-14     &  5.25-14   &  3.186-13     &	3.18-13   \\
   25   &   2s2p$^6$3s  	   &      $^1$S$  _{0}$   &  5.917-14     &  5.95-14   &  1.106-13     &  1.01-13   &  1.673-13     &	1.56-13   \\
   26   &   2s2p$^6$3p  	   &      $^3$P$^o_{1}$   &  1.070-15     &  9.92-16   &  4.911-14     &  4.94-14   &  3.846-14     &	3.87-14   \\
   27   &   2s2p$^6$3p  	   &      $^3$P$^o_{0}$   &  5.551-14     &  5.59-14   &  1.192-15     &  1.12-15   &  1.318-15     &	1.27-15   \\
   28   &   2s$^2$2p$^5$3d	   &      $^3$F$^o_{2}$   &  9.359-13     &  9.40-13   &  8.329-13     &  8.37-13   &  6.591-13     &	6.62-13   \\
   29   &   2s$^2$2p$^5$3d	   &      $^3$D$^o_{1}$   &  1.050-15     &  1.12-15   &  8.541-16     &  8.84-16   &  6.601-16     &	6.67-16   \\
   30   &   2s$^2$2p$^5$3d	   &      $^3$P$^o_{2}$   &  3.981-12     &  4.04-12   &  3.632-12     &  3.68-12   &  3.033-12     &	3.07-12   \\
   31   &   2s$^2$2p$^5$3d	   &      $^1$F$^o_{3}$   &  6.101-12     &  6.13-12   &  5.640-12     &  5.67-12   &  4.794-12     &	4.82-12   \\
   32   &   2s2p$^6$3p  	   &      $^3$P$^o_{2}$   &  5.165-14     &  5.19-14   &  4.568-14     &  4.59-14   &  3.573-14     &	3.59-14   \\
   33   &   2s2p$^6$3p  	   &      $^1$P$^o_{1}$   &  2.413-15     &  2.39-15   &  2.276-15     &  2.25-15   &  2.030-15     &	2.01-15   \\
   34   &   2s2p$^6$3d  	   &      $^3$D$  _{1}$   &  5.756-14     &  5.78-14   &  5.092-14     &  5.12-14   &  3.989-14     &	4.01-14   \\
   35   &   2s2p$^6$3d  	   &      $^3$D$  _{2}$   &  4.456-14     &  4.47-14   &  3.947-14     &  3.96-14   &  3.102-14     &	3.11-14   \\
   36   &   2s2p$^6$3d  	   &      $^3$D$  _{3}$   &  5.696-14     &  5.72-14   &  5.042-14     &  5.07-14   &  3.953-14     &	3.97-14   \\
   37   &   2s2p$^6$3d  	   &      $^1$D$  _{2}$   &  3.738-14     &  3.74-14   &  3.347-14     &  3.35-14   &  2.684-14     &	2.69-14   \\
   38   &   2s$^2$2p$^5$4s	   &      $^3$P$^o_{2}$   &  1.692-14     &  1.67-14   &  1.577-14     &  1.56-14   &  1.372-14     &	1.36-14   \\
   39   &   2s$^2$2p$^5$4s	   &      $^1$P$^o_{1}$   &  9.468-15     &  9.27-15   &  8.835-15     &  8.65-15   &  7.709-15     &	7.56-15   \\
   40   &   2s$^2$2p$^5$4p	   &      $^3$P$  _{1}$   &  1.393-14     &  1.38-14   &  1.299-14     &  1.29-14   &  1.133-14     &	1.12-14   \\
   41   &   2s$^2$2p$^5$4p	   &      $^3$D$  _{2}$   &  1.382-14     &  1.37-14   &  1.289-14     &  1.28-14   &  1.125-14     &	1.12-14   \\
   42   &   2s$^2$2p$^5$4p	   &      $^3$D$  _{3}$   &  2.128-14     &  2.10-14   &  2.015-14     &  1.99-14   &  1.813-14     &	1.79-14   \\
   43   &   2s$^2$2p$^5$4p	   &      $^1$P$  _{1}$   &  2.116-14     &  2.09-14   &  2.004-14     &  1.98-14   &  1.803-14     &	1.78-14   \\
   44   &   2s$^2$2p$^5$4p	   &      $^3$P$  _{2}$   &  2.139-14     &  2.11-14   &  2.024-14     &  2.00-14   &  1.820-14     &	1.80-14   \\
   45   &   2s$^2$2p$^5$4p	   &      $^1$S$  _{0}$   &  2.212-14     &  2.18-14   &  2.092-14     &  2.07-14   &  1.878-14     &	1.86-14   \\
   46   &   2s$^2$2p$^5$4d	   &      $^3$P$^o_{0}$   &  9.262-15     &  9.19-15   &  8.717-15     &  8.65-15   &  7.746-15     &	7.69-15   \\
   47   &   2s$^2$2p$^5$4d	   &      $^3$P$^o_{1}$   &  7.476-15     &  7.44-15   &  6.965-15     &  6.93-15   &  6.068-15     &	6.04-15   \\
   48   &   2s$^2$2p$^5$4d	   &      $^3$F$^o_{3}$   &  9.302-15     &  9.23-15   &  8.752-15     &  8.69-15   &  7.774-15     &	7.72-15   \\
   49   &   2s$^2$2p$^5$4d	   &      $^3$D$^o_{2}$   &  9.343-15     &  9.28-15   &  8.792-15     &  8.73-15   &  7.809-15     &	7.75-15   \\
   50   &   2s$^2$2p$^5$4d	   &      $^3$F$^o_{4}$   &  8.979-15     &  8.92-15   &  8.435-15     &  8.38-15   &  7.464-15     &	7.42-15   \\
\\  \hline            								                	 
\end{longtable}   								   					       
			      							   					       

														       
\begin{flushleft}													       
{\small
GRASP$^a$: Present calculations  with the GRASP code for 3948 levels \\
GRASP$^b$: Calculations of Singh et al. \cite{sam}  with the GRASP code for 3948 levels \\  
}															       
\end{flushleft} 

\clearpage
\newpage


\TableExplanation

\bigskip
\renewcommand{\arraystretch}{1.0}

\section*{Table 1.\label{tbl1te} Energies (Ryd) for the lowest 121 levels of Hf~LXIII and their lifetimes ($\tau$, s).}
\begin{tabular}{@{}p{1in}p{6in}@{}}
Index            & Level Index \\
Configuration    & The configuration to which the level belongs \\
Level             & The $LSJ$ designation of the level \\
GRASP$^a$          & Present energies from the GRASP code  with 64  configurations and 3948 level calculations \\
GRASP$^b$          & Earlier energies of Singh et al. \cite{sam}  from the GRASP code  with 64  configurations and 3948 level calculations \\
FAC               & Present energies from the FAC code  with 93~437 level calculations \\
$\tau$ (s)       & Lifetime of the level in s with the GRASP code \\

\end{tabular}
\label{tableII}

\bigskip
\renewcommand{\arraystretch}{1.0}

\section*{Table 2.\label{tbl2te} Energies (Ryd) for the lowest 121 levels of Ta~LXIV and their lifetimes ($\tau$, s).}
\begin{tabular}{@{}p{1in}p{6in}@{}}
Index            & Level Index \\
Configuration    & The configuration to which the level belongs \\
Level             & The $LSJ$ designation of the level \\
GRASP$^a$          & Present energies from the GRASP code  with 64  configurations and 3948 level calculations \\
GRASP$^b$          & Earlier energies of Singh et al. \cite{sam}  from the GRASP code  with 64  configurations and 3948 level calculations \\
FAC            & Present energies from the FAC code  with 93~437 level calculations \\
$\tau$ (s)       & Lifetime of the level in s \\

\end{tabular}
\label{tableII}

\bigskip
\renewcommand{\arraystretch}{1.0}

\section*{Table 3.\label{tbl3te} Energies (Ryd) for the lowest  117 levels of Re~LXVI and their lifetimes ($\tau$, s).}
\begin{tabular}{@{}p{1in}p{6in}@{}}
Index            & Level Index \\
Configuration    & The configuration to which the level belongs \\
Level             & The $LSJ$ designation of the level \\
GRASP$^a$          & Present energies from the GRASP code  with 64  configurations and 3948 level calculations \\
GRASP$^b$          & Earlier energies of Singh et al. \cite{sam}  from the GRASP code  with 64  configurations and 3948 level calculations \\
FAC               & Present energies from the FAC code  with 93~437  level calculations \\
$\tau$ (s)       & Lifetime of the level in s with the GRASP code \\

\end{tabular}
\label{tableII}

\bigskip
\renewcommand{\arraystretch}{1.0}


\bigskip
\section*{Table 4.\label{tbl4te}  Transition wavelengths ($\lambda_{ij}$ in $\rm \AA$), radiative rates (A$_{ji}$ in s$^{-1}$),
 oscillator strengths (f$_{ij}$, dimensionless), and line strengths (S, in atomic units) for electric dipole (E1), and 
A$_{ji}$ for electric quadrupole (E2), magnetic dipole (M1) and magnetic quadrupole (M2) transitions of Hf~LXIII.
 The ratio R(E1) of velocity and length forms of A-values for E1 transitions is listed in the last column.}
\begin{tabular}{@{}p{1in}p{6in}@{}}
$i$ and $j$         & The lower ($i$) and upper ($j$) levels of a transition as defined in Table 1.\\
$\lambda_{ij}$      & Transition wavelength (in ${\rm \AA}$) \\
A$^{E1}_{ji}$       & Radiative transition probability (in s$^{-1}$) for the E1 transitions \\
f$^{E1}_{ij}$       & Absorption oscillator strength (dimensionless) for the E1 transitions \\
S$^{E1}$            & Line strength in atomic unit (a.u.), 1 a.u. = 6.460$\times$10$^{-36}$ cm$^2$ esu$^2$ for the E1 transitions \\
A$^{E2}_{ji}$       & Radiative transition probability (in s$^{-1}$) for the E2 transitions \\
A$^{M1}_{ji}$       & Radiative transition probability (in s$^{-1}$) for the M1 transitions \\
A$^{M2}_{ji}$       & Radiative transition probability (in s$^{-1}$) for the M2 transitions \\
R(E1)                     & Ratio of velocity and length forms of A- (or f- and S-) values for the E1 transitions \\
$a{\pm}b$ &  $\equiv a\times{10^{{\pm}b}}$ \\
\end{tabular}
\label{ExplTable4}

\bigskip
\section*{Table 5.\label{tbl5te}  Transition wavelengths ($\lambda_{ij}$ in $\rm \AA$), radiative rates (A$_{ji}$ in s$^{-1}$),
 oscillator strengths (f$_{ij}$, dimensionless), and line strengths (S, in atomic units) for electric dipole (E1), and 
A$_{ji}$ for electric quadrupole (E2), magnetic dipole (M1) and magnetic quadrupole (M2) transitions of Ta~LXIV.
The ratio R(E1) of velocity and length forms of A-values for E1 transitions is listed in the last column.}
\begin{tabular}{@{}p{1in}p{6in}@{}}
$i$ and $j$         & The lower ($i$) and upper ($j$) levels of a transition as defined in Table 2.\\
$\lambda_{ij}$      & Transition wavelength (in ${\rm \AA}$) \\
A$^{E1}_{ji}$       & Radiative transition probability (in s$^{-1}$) for the E1 transitions \\
f$^{E1}_{ij}$       & Absorption oscillator strength (dimensionless) for the E1 transitions \\
S$^{E1}$            & Line strength in atomic unit (a.u.), 1 a.u. = 6.460$\times$10$^{-36}$ cm$^2$ esu$^2$ for the E1 transitions \\
A$^{E2}_{ji}$       & Radiative transition probability (in s$^{-1}$) for the E2 transitions \\
A$^{M1}_{ji}$       & Radiative transition probability (in s$^{-1}$) for the M1 transitions \\
A$^{M2}_{ji}$       & Radiative transition probability (in s$^{-1}$) for the M2 transitions \\
R(E1)                     & Ratio of velocity and length forms of A- (or f- and S-) values for the E1 transitions \\
$a{\pm}b$ &  $\equiv a\times{10^{{\pm}b}}$ \\
\end{tabular}
\label{ExplTable5}

\bigskip
\section*{Table 6.\label{tbl6te}  Transition wavelengths ($\lambda_{ij}$ in $\rm \AA$), radiative rates (A$_{ji}$ in s$^{-1}$),
 oscillator strengths (f$_{ij}$, dimensionless), and line strengths (S, in atomic units) for electric dipole (E1), and 
A$_{ji}$ for electric quadrupole (E2), magnetic dipole (M1) and magnetic quadrupole (M2) transitions of Re~LXVI.
 The ratio R(E1) of velocity and length forms of A-values for E1 transitions is listed in the last column.}
\begin{tabular}{@{}p{1in}p{6in}@{}}
$i$ and $j$         & The lower ($i$) and upper ($j$) levels of a transition as defined in Table 3.\\
$\lambda_{ij}$      & Transition wavelength (in ${\rm \AA}$) \\
A$^{E1}_{ji}$       & Radiative transition probability (in s$^{-1}$) for the E1 transitions \\
f$^{E1}_{ij}$       & Absorption oscillator strength (dimensionless) for the E1 transitions \\
S$^{E1}$            & Line strength in atomic unit (a.u.), 1 a.u. = 6.460$\times$10$^{-36}$ cm$^2$ esu$^2$ for the E1 transitions \\
A$^{E2}_{ji}$       & Radiative transition probability (in s$^{-1}$) for the E2 transitions \\
A$^{M1}_{ji}$       & Radiative transition probability (in s$^{-1}$) for the M1 transitions \\
A$^{M2}_{ji}$       & Radiative transition probability (in s$^{-1}$) for the M2 transitions \\
R(E1)                     & Ratio of velocity and length forms of A- (or f- and S-) values for the E1 transitions \\
$a{\pm}b$ &  $\equiv a\times{10^{{\pm}b}}$ \\
\end{tabular}
\label{ExplTable6}


\datatables 



\setlength{\LTleft}{0pt}
\setlength{\LTright}{0pt} 


\setlength{\tabcolsep}{0.5\tabcolsep}

\renewcommand{\arraystretch}{1.0}

\footnotesize 

\begin{longtable}{@{\extracolsep\fill}rllrrrr@{}}
\caption{Energies (Ryd) for  the lowest  121 levels of Hf~LXIII and their lifetimes ($\tau$, s).  $a{\pm}b \equiv a{\times}$10$^{{\pm}b}$. See page\ \pageref{tbl1te} for Explanation of Tables} 
Index  &     Configuration          & Level     & GRASP$^a$   & GRASP$^b$   &   FAC   &  $\tau$ (s)   \\  \\
\hline\\
\endfirsthead\\
\caption[]{(continued)}
Index  &     Configuration          & Level     & GRASP$^a$   & GRASP$^b$   &   FAC   &  $\tau$ (s)  \\  \\
\hline\\
\endhead
    1   &   2s$^2$2p$^6$	   &      $^1$S$  _{0}$   &  	  0.0000   &     0.0000   &  	  0.0000   &     ........   \\
    2   &   2s$^2$2p$^5$3s	   &      $^3$P$^o_{2}$   &  	575.7815   &   575.7567   &  	575.8687   &     3.023-10   \\   
    3   &   2s$^2$2p$^5$3s	   &      $^1$P$^o_{1}$   &  	576.3806   &   576.3433   &  	576.4839   &     7.476-15   \\   
    4   &   2s$^2$2p$^5$3p	   &      $^3$P$  _{1}$   &  	586.3673   &   586.3632   &  	586.3682   &     1.823-11   \\ 
    5   &   2s$^2$2p$^5$3p	   &      $^3$D$  _{2}$   &  	586.5020   &   586.4994   &  	586.5079   &     1.238-12   \\ 
    6   &   2s$^2$2p$^5$3p	   &      $^1$P$  _{1}$   &  	610.8609   &   610.8442   &  	610.8692   &     8.091-13   \\
    7   &   2s$^2$2p$^5$3p	   &      $^3$D$  _{3}$   &  	610.8332   &   610.8157   &  	610.8431   &     7.685-13   \\
    8   &   2s$^2$2p$^5$3p	   &      $^3$P$  _{2}$   &  	611.7520   &   611.7430   &  	611.7739   &     4.653-13   \\
    9   &   2s$^2$2p$^5$3p	   &      $^1$S$  _{0}$   &  	615.8231   &   615.8706   &  	615.8828   &     5.139-13   \\
   10   &   2s$^2$2p$^5$3d	   &      $^3$P$^o_{0}$   &  	622.1597   &   622.1318   &  	622.0878   &     1.032-12   \\
   11   &   2s$^2$2p$^5$3d	   &      $^3$P$^o_{1}$   &  	622.8653   &   622.8380   &  	622.7967   &     1.538-14   \\
   12   &   2s$^2$2p$^5$3d	   &      $^3$F$^o_{3}$   &  	622.9534   &   622.9268   &  	622.8892   &     9.772-13   \\
   13   &   2s$^2$2p$^5$3d	   &      $^3$D$^o_{2}$   &  	623.5171   &   623.4919   &  	623.4488   &     9.292-13   \\
   14   &   2s$^2$2p$^5$3d	   &      $^3$F$^o_{4}$   &  	628.3074   &   628.2736   &  	628.2408   &     7.266-12   \\
   15   &   2s$^2$2p$^5$3d	   &      $^1$D$^o_{2}$   &  	628.8085   &   628.7769   &  	628.7407   &     4.178-12   \\
   16   &   2s$^2$2p$^5$3d	   &      $^3$D$^o_{3}$   &  	629.5257   &   629.4971   &  	629.4604   &     6.615-12   \\ 
   17   &   2s$^2$2p$^5$3d	   &      $^1$P$^o_{1}$   &  	631.6257   &   631.6176   &  	631.5499   &     4.052-16   \\ 
   18   &   2s$^2$2p$^5$3s	   &      $^3$P$^o_{0}$   &  	665.5676   &   665.5019   &  	665.7006   &     4.484-11   \\ 
   19   &   2s$^2$2p$^5$3s	   &      $^3$P$^o_{1}$   &  	665.8152   &   665.7470   &  	665.9550   &     3.059-14   \\ 
   20   &   2s$^2$2p$^5$3p	   &      $^3$D$  _{1}$   &  	675.9010   &   675.8588   &  	675.9448   &     1.844-11   \\ 
   21   &   2s$^2$2p$^5$3p	   &      $^3$P$  _{0}$   &  	679.3429   &   679.3396   &  	679.4135   &     3.717-12   \\ 
   22   &   2s2p$^6$3s  	   &      $^3$S$  _{1}$   &  	698.8042   &   698.9230   &  	698.8655   &     1.076-13   \\ 
   23   &   2s$^2$2p$^5$3p	   &      $^1$D$  _{2}$   &  	700.9383   &   700.8899   &  	700.9984   &     4.445-13   \\ 
   24   &   2s$^2$2p$^5$3p	   &      $^3$S$  _{1}$   &  	702.1490   &   702.2285   &  	702.0912   &     9.136-14   \\ 
   25   &   2s2p$^6$3s  	   &      $^1$S$  _{0}$   &  	702.4803   &   702.8162   &  	702.4548   &     5.917-14   \\ 
   26   &   2s2p$^6$3p  	   &      $^3$P$^o_{1}$   &  	710.5700   &   710.7913   &  	710.5004   &     1.070-15   \\ 
   27   &   2s2p$^6$3p  	   &      $^3$P$^o_{0}$   &  	710.8421   &   711.1017   &  	710.7154   &     5.551-14   \\ 
   28   &   2s$^2$2p$^5$3d	   &      $^3$F$^o_{2}$   &  	712.6998   &   712.6381   &  	712.6841   &     9.359-13   \\
   29   &   2s$^2$2p$^5$3d	   &      $^3$D$^o_{1}$   &  	714.7552   &   714.7445   &  	714.6761   &     1.050-15   \\
   30   &   2s$^2$2p$^5$3d	   &      $^3$P$^o_{2}$   &  	718.5311   &   718.4675   &  	718.5085   &     3.981-12   \\ 
   31   &   2s$^2$2p$^5$3d	   &      $^1$F$^o_{3}$   &  	718.7807   &   718.7166   &  	718.7590   &     6.101-12   \\ 
   32   &   2s2p$^6$3p  	   &      $^3$P$^o_{2}$   &  	735.4475   &   735.6877   &  	735.3331   &     5.165-14   \\ 
   33   &   2s2p$^6$3p  	   &      $^1$P$^o_{1}$   &  	735.8760   &   736.1144   &  	735.7688   &     2.413-15   \\ 
   34   &   2s2p$^6$3d  	   &      $^3$D$  _{1}$   &  	747.1808   &   747.4588   &  	746.9636   &     5.756-14   \\
   35   &   2s2p$^6$3d  	   &      $^3$D$  _{2}$   &  	747.5628   &   747.8408   &  	747.3556   &     4.456-14   \\
   36   &   2s2p$^6$3d  	   &      $^3$D$  _{3}$   &  	752.8109   &   753.0779   &  	752.5972   &     5.696-14   \\
   37   &   2s2p$^6$3d  	   &      $^1$D$  _{2}$   &  	753.9151   &   754.1868   &  	753.7184   &     3.738-14   \\
   38   &   2s$^2$2p$^5$4s	   &      $^3$P$^o_{2}$   &  	804.5593   &   804.5481   &  	804.4619   &     1.692-14   \\ 
   39   &   2s$^2$2p$^5$4s	   &      $^1$P$^o_{1}$   &  	804.7585   &   804.7387   &  	804.6636   &     9.468-15   \\ 
   40   &   2s$^2$2p$^5$4p	   &      $^3$P$  _{1}$   &  	808.9378   &   808.9345   &  	808.8040   &     1.393-14   \\
   41   &   2s$^2$2p$^5$4p	   &      $^3$D$  _{2}$   &  	808.9770   &   808.9740   &  	808.8440   &     1.382-14   \\
   42   &   2s$^2$2p$^5$4p	   &      $^3$D$  _{3}$   &  	818.9843   &   818.9747   &  	818.8565   &     2.128-14   \\
   43   &   2s$^2$2p$^5$4p	   &      $^1$P$  _{1}$   &  	819.0065   &   818.9972   &  	818.8774   &     2.116-14   \\
   44   &   2s$^2$2p$^5$4p	   &      $^3$P$  _{2}$   &  	819.3236   &   819.3168   &  	819.1969   &     2.139-14   \\
   45   &   2s$^2$2p$^5$4p	   &      $^1$S$  _{0}$   &  	820.6703   &   820.7220   &  	820.5433   &     2.212-14   \\ 
   46   &   2s$^2$2p$^5$4d	   &      $^3$P$^o_{0}$   &  	823.3211   &   823.3130   &  	823.1661   &     9.262-15   \\ 
   47   &   2s$^2$2p$^5$4d	   &      $^3$P$^o_{1}$   &  	823.5749   &   823.5683   &  	823.4191   &     7.476-15   \\
   48   &   2s$^2$2p$^5$4d	   &      $^3$F$^o_{3}$   &  	823.5919   &   823.5860   &  	823.4349   &     9.302-15   \\ 
   49   &   2s$^2$2p$^5$4d	   &      $^3$D$^o_{2}$   &  	823.7955   &   823.7916   &  	823.6368   &     9.343-15   \\ 
   50   &   2s$^2$2p$^5$4d	   &      $^3$F$^o_{4}$   &  	825.9321   &   825.9214   &  	825.7770   &     8.979-15   \\ 
   51   &   2s$^2$2p$^5$4d	   &      $^1$D$^o_{2}$   &  	826.1163   &   826.1076   &  	825.9588   &     8.979-15   \\ 
   52   &   2s$^2$2p$^5$4d	   &      $^3$D$^o_{3}$   &  	826.3718   &   826.3659   &  	826.2126   &     9.034-15   \\ 
   53   &   2s$^2$2p$^5$4d	   &      $^1$P$^o_{1}$   &  	827.1047   &   827.1090   &  	826.9314   &     1.003-15   \\
   54   &   2s$^2$2p$^5$4f	   &      $^3$D$  _{1}$   &  	828.4688   &   828.4424   &  	828.3033   &     4.264-15   \\ 
   55   &   2s$^2$2p$^5$4f	   &      $^3$G$  _{4}$   &  	828.4918   &   828.4654   &  	828.3289   &     4.371-15   \\ 
   56   &   2s$^2$2p$^5$4f	   &      $^3$D$  _{2}$   &  	828.6391   &   828.6134   &  	828.4755   &     4.310-15   \\
   57   &   2s$^2$2p$^5$4f	   &      $^3$F$  _{3}$   &  	828.7164   &   828.6907   &  	828.5534   &     4.364-15   \\
   58   &   2s$^2$2p$^5$4f	   &      $^3$G$  _{5}$   &  	829.5781   &   829.5507   &  	829.4128   &     4.474-15   \\ 
   59   &   2s$^2$2p$^5$4f	   &      $^1$D$  _{2}$   &  	829.6955   &   829.6708   &  	829.5327   &     4.362-15   \\
   60   &   2s$^2$2p$^5$4f	   &      $^1$F$  _{3}$   &  	829.7577   &   829.7313   &  	829.5917   &     4.438-15   \\
   61   &   2s$^2$2p$^5$4f	   &      $^3$F$  _{4}$   &  	829.8270   &   829.8009   &  	829.6628   &     4.486-15   \\
   62   &   2s$^2$2p$^5$4s	   &      $^3$P$^o_{0}$   &  	894.4662   &   894.4119   &  	894.3708   &     1.672-14   \\ 
   63   &   2s$^2$2p$^5$4s	   &      $^3$P$^o_{1}$   &  	894.5350   &   894.4809   &  	894.4406   &     1.312-14   \\ 
   64   &   2s$^2$2p$^5$4p	   &      $^3$D$  _{1}$   &  	898.7422   &   898.6946   &  	898.6089   &     1.374-14   \\ 
   65   &   2s$^2$2p$^5$4p	   &      $^3$P$  _{0}$   &  	899.8223   &   899.8053   &  	899.6935   &     1.420-14   \\ 
   66   &   2s$^2$2p$^5$5s	   &      $^3$P$^o_{2}$   &  	906.6111   &   906.5924   &  	906.4735   &     2.117-14   \\
   67   &   2s$^2$2p$^5$5s	   &      $^1$P$^o_{1}$   &  	906.7035   &   906.6865   &  	906.5666   &     1.313-14   \\
   68   &   2s$^2$2p$^5$5p	   &      $^3$P$  _{1}$   &  	908.6592   &   908.6230   &  	908.5558   &     1.892-14   \\
   69   &   2s$^2$2p$^5$5p	   &      $^3$D$  _{2}$   &  	908.7659   &   908.7339   &  	908.6430   &     1.873-14   \\
   70   &   2s$^2$2p$^5$4p	   &      $^3$S$  _{1}$   &  	909.1010   &   909.0720   &  	908.9217   &     1.954-14   \\
   71   &   2s$^2$2p$^5$4p	   &      $^1$D$  _{2}$   &  	909.0650   &   909.0320   &  	908.9061   &     1.949-14   \\
   72   &   2s$^2$2p$^5$5p	   &      $^3$D$  _{3}$   &  	913.8747   &   913.8570   &  	913.7206   &     2.552-14   \\
   73   &   2s$^2$2p$^5$5p	   &      $^1$P$  _{1}$   &  	913.8848   &   913.8675   &  	913.7300   &     2.497-14   \\
   74   &   2s$^2$2p$^5$5p	   &      $^3$P$  _{2}$   &  	914.0394   &   914.0248   &  	913.8853   &     2.567-14   \\
   75   &   2s$^2$2p$^5$4d	   &      $^3$F$^o_{2}$   &  	913.4705   &   913.4202   &  	913.3179   &     9.309-15   \\
   76   &   2s$^2$2p$^5$5p	   &      $^1$S$  _{0}$   &  	914.6780   &   914.6979   &  	914.5204   &     2.633-14   \\ 
   77   &   2s$^2$2p$^5$4d	   &      $^3$D$^o_{1}$   &  	914.0253   &   913.9820   &  	913.8671   &     1.809-15   \\ 
   78   &   2s$^2$2p$^5$5d	   &      $^3$P$^o_{0}$   &  	916.0073   &   915.9891   &  	915.8451   &     1.253-14   \\ 
   79   &   2s$^2$2p$^5$5d	   &      $^3$P$^o_{1}$   &  	916.1315   &   916.1151   &  	915.9681   &     1.121-14   \\ 
   80   &   2s$^2$2p$^5$5d	   &      $^3$F$^o_{3}$   &  	916.1925   &   916.1712   &  	915.8375   &     1.170-14   \\ 
   81   &   2s$^2$2p$^5$5d	   &      $^3$D$^o_{2}$   &  	916.2616   &   916.2462   &  	915.7951   &     1.239-14   \\ 
   82   &   2s$^2$2p$^5$4d	   &      $^3$P$^o_{2}$   &  	915.9458   &   915.8954   &  	916.1042   &     9.116-15   \\ 
   83   &   2s$^2$2p$^5$4d	   &      $^1$F$^o_{3}$   &  	916.0225   &   915.9763   &  	916.0604   &     9.563-15   \\ 
   84   &   2s$^2$2p$^5$5d	   &      $^3$F$^o_{4}$   &  	917.3463   &   917.3267   &  	917.1839   &     1.251-14   \\ 
   85   &   2s$^2$2p$^5$5d	   &      $^1$D$^o_{2}$   &  	917.4647   &   917.4464   &  	917.2903   &     1.242-14   \\ 
   86   &   2s$^2$2p$^5$5d	   &      $^3$D$^o_{3}$   &  	917.5569   &   917.5415   &  	917.3904   &     1.253-14   \\ 
   87   &   2s$^2$2p$^5$5d	   &      $^1$P$^o_{1}$   &  	917.9588   &   917.9341   &  	917.7794   &     1.631-15   \\ 
   88   &   2s$^2$2p$^5$5f	   &      $^3$D$  _{1}$   &  	918.5828   &   918.5560   &  	918.4185   &     8.224-15   \\
   89   &   2s$^2$2p$^5$5f	   &      $^3$G$  _{4}$   &  	918.6071   &   918.5807   &  	918.4443   &     8.357-15   \\
   90   &   2s$^2$2p$^5$5f	   &      $^3$D$  _{2}$   &  	918.7023   &   918.6728   &  	918.3915   &     7.471-15   \\ 
   91   &   2s$^2$2p$^5$5f	   &      $^3$F$  _{3}$   &  	918.7228   &   918.6971   &  	918.3610   &     8.071-15   \\ 
   92   &   2s$^2$2p$^5$5f	   &      $^3$G$  _{5}$   &  	919.1627   &   919.1355   &  	918.9985   &     8.577-15   \\ 
   93   &   2s$^2$2p$^5$5f	   &      $^1$D$  _{2}$   &  	919.2361   &   919.2113   &  	918.5605   &     8.337-15   \\ 
   94   &   2s$^2$2p$^5$5f	   &      $^1$F$  _{3}$   &  	919.2257   &   919.1962   &  	918.5645   &     7.995-15   \\
   95   &   2s$^2$2p$^5$5f	   &      $^3$F$  _{4}$   &  	919.2796   &   919.2527   &  	919.1179   &     8.343-15   \\
   96   &   2s$^2$2p$^5$5g	   &      $^3$F$^o_{2}$   &  	919.3577   &   919.3272   &  	919.0732   &     1.466-14   \\
   97   &   2s$^2$2p$^5$5g	   &      $^3$H$^o_{5}$   &  	919.3642   &   919.3336   &  	919.1828   &     1.458-14   \\
   98   &   2s$^2$2p$^5$5g	   &      $^3$F$^o_{3}$   &  	919.4094   &   919.3789   &  	919.0651   &     1.463-14   \\ 
   99   &   2s$^2$2p$^5$5g	   &      $^3$G$^o_{4}$   &  	919.4283   &   919.3977   &  	919.2473   &     1.459-14   \\ 
  100   &   2s$^2$2p$^5$4f	   &      $^3$G$  _{3}$   &  	918.5253   &   918.4568   &  	919.2285   &     4.434-15   \\
  101   &   2s$^2$2p$^5$4f	   &      $^3$F$  _{2}$   &  	918.5721   &   918.5080   &  	919.1767   &     4.550-15   \\
  102   &   2s$^2$2p$^5$5g	   &      $^1$F$^o_{3}$   &  	919.7008   &   919.6703   &  	919.5196   &     1.484-14   \\
  103   &   2s$^2$2p$^5$5g	   &      $^3$H$^o_{6}$   &  	919.7029   &   919.6722   &  	919.5213   &     1.474-14   \\
  104   &   2s$^2$2p$^5$5g	   &      $^1$G$^o_{4}$   &  	919.7503   &   919.7197   &  	919.5058   &     1.479-14   \\
  105   &   2s$^2$2p$^5$5g	   &      $^3$G$^o_{5}$   &  	919.7657   &   919.7351   &  	919.5847   &     1.476-14   \\ 
  106   &   2s$^2$2p$^5$4f	   &      $^3$D$  _{3}$   &  	919.6913   &   919.6257   &  	919.5233   &     4.604-15   \\ 
  107   &   2s$^2$2p$^5$4f	   &      $^1$G$  _{4}$   &  	919.6703   &   919.6026   &  	919.5692   &     4.549-15   \\
  108   &   2s2p$^6$4s  	   &      $^3$S$  _{1}$   &  	928.7387   &   928.9314   &  	928.5050   &     1.299-14   \\ 
  109   &   2s2p$^6$4s  	   &      $^1$S$  _{0}$   &  	929.4870   &   929.7128   &  	929.2485   &     1.308-14   \\ 
  110   &   2s2p$^6$4p  	   &      $^3$P$^o_{0}$   &  	933.1438   &   933.3486   &  	932.8751   &     1.123-14   \\ 
  111   &   2s2p$^6$4p  	   &      $^3$P$^o_{1}$   &  	933.1757   &   933.3806   &  	932.9070   &     4.710-15   \\ 
  112   &   2s2p$^6$4p  	   &      $^3$P$^o_{2}$   &  	943.2514   &   943.4473   &  	942.9907   &     1.558-14   \\ 
  113   &   2s2p$^6$4p  	   &      $^1$P$^o_{1}$   &  	943.4009   &   943.5986   &  	943.1393   &     3.693-15   \\
  114   &   2s2p$^6$4d  	   &      $^3$D$  _{1}$   &  	947.7445   &   947.9465   &  	947.4497   &     7.968-15   \\ 
  115   &   2s2p$^6$4d  	   &      $^3$D$  _{2}$   &  	947.8770   &   948.0809   &  	947.5840   &     7.931-15   \\ 
  116   &   2s2p$^6$4d  	   &      $^3$D$  _{3}$   &  	950.1801   &   950.3751   &  	949.8898   &     7.724-15   \\
  117   &   2s2p$^6$4d  	   &      $^1$D$  _{2}$   &  	950.5599   &   950.7632   &  	950.2650   &     7.647-15   \\
  118   &   2s2p$^6$4f  	   &      $^3$F$^o_{2}$   &  	952.6857   &   952.8481   &  	952.3794   &     4.022-15   \\ 
  119   &   2s2p$^6$4f  	   &      $^3$F$^o_{3}$   &  	952.6693   &   952.8314   &  	952.3687   &     4.030-15   \\
  120   &   2s2p$^6$4f  	   &      $^3$F$^o_{4}$   &  	953.7491   &   953.9108   &  	953.4471   &     4.120-15   \\
  121   &   2s2p$^6$4f  	   &      $^1$F$^o_{3}$   &  	953.8320   &   953.9939   &  	953.5240   &     4.135-15   \\
   \hline  											      
			      							   					       

\\
\end{longtable}
\clearpage
\newpage

\begin{longtable}{@{\extracolsep\fill}rllrrrrr@{}}
\caption{Energies (Ryd) for   the lowest  121 levels of Ta~LXIV and their lifetimes ($\tau$, s).   $a{\pm}b \equiv a{\times}$10$^{{\pm}b}$. See page\ \pageref{tbl2te} for Explanation of Tables} 
Index  &     Configuration                  & Level    & GRASP$^a$   & GRASP$^b$   &   FAC   &  $\tau$ (s)  \\  \\
\hline\\
\endfirsthead\\
\caption[]{(continued)}
Index  &     Configuration                  & Level     & GRASP$^a$   & GRASP$^b$   &   FAC   &  $\tau$ (s)   \\  \\
\hline\\
\endhead
    1   &   2s$^2$2p$^6$	   &      $^1$S$  _{0}$    &	  0.0000   &	   0.0000    &       0.0000   &    ........   \\
    2   &   2s$^2$2p$^5$3s	   &      $^3$P$^o_{2}$    &	592.6332   &	 592.5993    &     592.7304   &    2.656-10   \\
    3   &   2s$^2$2p$^5$3s	   &      $^1$P$^o_{1}$    &	593.2441   &	 593.2078    &     593.3575   &    6.988-15   \\
    4   &   2s$^2$2p$^5$3p	   &      $^3$P$  _{1}$    &	603.4590   &	 603.4562    &     603.4594   &    1.678-11   \\
    5   &   2s$^2$2p$^5$3p	   &      $^3$D$  _{2}$    &	603.5900   &	 603.5888    &     603.5955   &    1.136-12   \\
    6   &   2s$^2$2p$^5$3p	   &      $^1$P$  _{1}$    &	629.5979   &	 629.5819    &     629.6072   &    7.131-13   \\
    7   &   2s$^2$2p$^5$3p	   &      $^3$D$  _{3}$    &	629.5679   &	 629.5512    &     629.5788   &    6.788-13   \\
    8   &   2s$^2$2p$^5$3p	   &      $^3$P$  _{2}$    &	630.5058   &	 630.4977    &     630.5287   &    4.166-13   \\
    9   &   2s$^2$2p$^5$3p	   &      $^1$S$  _{0}$    &	634.6600   &	 634.7089    &     634.7201   &    4.593-13   \\
   10   &   2s$^2$2p$^5$3d	   &      $^3$P$^o_{0}$    &	641.0972   &	 641.0699    &     641.0226   &    9.172-13   \\
   11   &   2s$^2$2p$^5$3d	   &      $^3$P$^o_{1}$    &	641.8166   &	 641.7900    &     641.7455   &    1.359-14   \\
   12   &   2s$^2$2p$^5$3d	   &      $^3$F$^o_{3}$    &	641.8922   &	 641.8662    &     641.8254   &    8.707-13   \\
   13   &   2s$^2$2p$^5$3d	   &      $^3$D$^o_{2}$    &	642.4723   &	 642.4478    &     642.4015   &    8.288-13   \\
   14   &   2s$^2$2p$^5$3d	   &      $^3$F$^o_{4}$    &	647.6156   &	 647.5823    &     647.5465   &    6.805-12   \\
   15   &   2s$^2$2p$^5$3d	   &      $^1$D$^o_{2}$    &	648.1290   &	 648.0979    &     648.0587   &    3.831-12   \\
   16   &   2s$^2$2p$^5$3d	   &      $^3$D$^o_{3}$    &	648.8576   &	 648.8295    &     648.7899   &    6.205-12   \\
   17   &   2s$^2$2p$^5$3d	   &      $^1$P$^o_{1}$    &	650.9870   &	 650.9793    &     650.9077   &    3.818-16   \\
   18   &   2s$^2$2p$^5$3s	   &      $^3$P$^o_{0}$    &	688.3005   &	 688.2345    &     688.4484   &    3.743-11   \\
   19   &   2s$^2$2p$^5$3s	   &      $^3$P$^o_{1}$    &	688.5490   &	 688.4805    &     688.7037   &    2.816-14   \\
   20   &   2s$^2$2p$^5$3p	   &      $^3$D$  _{1}$    &	698.8569   &	 698.8145    &     698.9050   &    1.679-11   \\
   21   &   2s$^2$2p$^5$3p	   &      $^3$P$  _{0}$    &	702.3737   &	 702.3704    &     702.4484   &    3.435-12   \\
   22   &   2s2p$^6$3s  	   &      $^3$S$  _{1}$    &	722.6310   &	 722.7934    &     722.6465   &    7.255-14   \\
   23   &   2s$^2$2p$^5$3p	   &      $^1$D$  _{2}$    &	725.5578   &	 725.5080    &     725.6237   &    3.977-13   \\
   24   &   2s2p$^6$3s  	   &      $^1$S$  _{0}$    &	726.0245   &	 726.3644    &     725.9777   &    5.224-14   \\
   25   &   2s$^2$2p$^5$3p	   &      $^3$S$  _{1}$    &	726.4308   &	 726.4693    &     726.4039   &    1.106-13   \\
   26   &   2s2p$^6$3p  	   &      $^3$P$^o_{0}$    &	734.5706   &	 734.8333    &     734.4127   &    4.911-14   \\
   27   &   2s2p$^6$3p  	   &      $^3$P$^o_{1}$    &	734.3654   &	 734.6042    &     734.2450   &    1.192-15   \\
   28   &   2s$^2$2p$^5$3d	   &      $^3$F$^o_{2}$    &	737.5238   &	 737.4615    &     737.5106   &    8.329-13   \\
   29   &   2s$^2$2p$^5$3d	   &      $^3$D$^o_{1}$    &	739.5257   &	 739.5000    &     739.4681   &    8.541-16   \\
   30   &   2s$^2$2p$^5$3d	   &      $^3$P$^o_{2}$    &	743.7347   &	 743.6703    &     743.7145   &    3.632-12   \\
   31   &   2s$^2$2p$^5$3d	   &      $^1$F$^o_{3}$    &	743.9855   &	 743.9205    &     743.9662   &    5.640-12   \\
   32   &   2s2p$^6$3p  	   &      $^3$P$^o_{2}$    &	760.8252   &	 761.0678    &     760.6813   &    4.568-14   \\
   33   &   2s2p$^6$3p  	   &      $^1$P$^o_{1}$    &	761.2615   &	 761.5023    &     761.1247   &    2.276-15   \\
   34   &   2s2p$^6$3d  	   &      $^3$D$  _{1}$    &	772.7641   &	 773.0447    &     772.5134   &    5.092-14   \\
   35   &   2s2p$^6$3d  	   &      $^3$D$  _{2}$    &	773.1502   &	 773.4308    &     772.9099   &    3.947-14   \\
   36   &   2s2p$^6$3d  	   &      $^3$D$  _{3}$    &	778.7634   &	 779.0326    &     778.5166   &    5.042-14   \\
   37   &   2s2p$^6$3d  	   &      $^1$D$  _{2}$    &	779.8813   &	 780.1552    &     779.6514   &    3.347-14   \\
   38   &   2s$^2$2p$^5$4s	   &      $^3$P$^o_{2}$    &	829.1310   &	 829.1212    &     829.0386   &    1.577-14   \\
   39   &   2s$^2$2p$^5$4s	   &      $^1$P$^o_{1}$    &	829.3340   &	 829.3256    &     829.2440   &    8.835-15   \\
   40   &   2s$^2$2p$^5$4p	   &      $^3$P$  _{1}$    &	833.6067   &	 833.6049    &     833.4734   &    1.299-14   \\
   41   &   2s$^2$2p$^5$4p	   &      $^3$D$  _{2}$    &	833.6447   &	 833.6431    &     833.5122   &    1.289-14   \\
   42   &   2s$^2$2p$^5$4p	   &      $^3$D$  _{3}$    &	844.3295   &	 844.3212    &     844.2029   &    2.015-14   \\
   43   &   2s$^2$2p$^5$4p	   &      $^1$P$  _{1}$    &	844.3526   &	 844.3446    &     844.2247   &    2.004-14   \\
   44   &   2s$^2$2p$^5$4p	   &      $^3$P$  _{2}$    &	844.6760   &	 844.6705    &     844.5505   &    2.024-14   \\
   45   &   2s$^2$2p$^5$4p	   &      $^1$S$  _{0}$    &	846.0486   &	 846.1021    &     845.9225   &    2.092-14   \\
   46   &   2s$^2$2p$^5$4d	   &      $^3$P$^o_{0}$    &	848.7431   &	 848.7364    &     848.5876   &    8.717-15   \\
   47   &   2s$^2$2p$^5$4d	   &      $^3$P$^o_{1}$    &	849.0016   &	 848.9964    &     848.8453   &    6.965-15   \\
   48   &   2s$^2$2p$^5$4d	   &      $^3$F$^o_{3}$    &	849.0143   &	 849.0097    &     848.8569   &    8.752-15   \\
   49   &   2s$^2$2p$^5$4d	   &      $^3$D$^o_{2}$    &	849.2240   &	 849.2215    &     849.0648   &    8.792-15   \\
   50   &   2s$^2$2p$^5$4d	   &      $^3$F$^o_{4}$    &	851.5140   &	 851.5045    &     851.3585   &    8.435-15   \\
   51   &   2s$^2$2p$^5$4d	   &      $^1$D$^o_{2}$    &	851.7028   &	 851.6954    &     851.5449   &    8.434-15   \\
   52   &   2s$^2$2p$^5$4d	   &      $^3$D$^o_{3}$    &	851.9626   &	 851.9580    &     851.8031   &    8.485-15   \\
   53   &   2s$^2$2p$^5$4d	   &      $^1$P$^o_{1}$    &	852.7055   &	 852.7110    &     852.5315   &    9.462-16   \\
   54   &   2s$^2$2p$^5$4f	   &      $^3$D$  _{1}$    &	854.0911   &	 854.0658    &     853.9250   &    4.001-15   \\
   55   &   2s$^2$2p$^5$4f	   &      $^3$G$  _{4}$    &	854.1117   &	 854.0864    &     853.9484   &    4.102-15   \\
   56   &   2s$^2$2p$^5$4f	   &      $^3$D$  _{2}$    &	854.2639   &	 854.2393    &     854.0997   &    4.045-15   \\
   57   &   2s$^2$2p$^5$4f	   &      $^3$F$  _{3}$    &	854.3415   &	 854.3171    &     854.1781   &    4.096-15   \\
   58   &   2s$^2$2p$^5$4f	   &      $^3$G$  _{5}$    &	855.2706   &	 855.2444    &     855.1048   &    4.204-15   \\
   59   &   2s$^2$2p$^5$4f	   &      $^1$D$  _{2}$    &	855.3919   &	 855.3683    &     855.2285   &    4.093-15   \\
   60   &   2s$^2$2p$^5$4f	   &      $^1$F$  _{3}$    &	855.4548   &	 855.4295    &     855.2881   &    4.169-15   \\
   61   &   2s$^2$2p$^5$4f	   &      $^3$F$  _{4}$    &	855.5246   &	 855.4997    &     855.3599   &    4.215-15   \\
   62   &   2s$^2$2p$^5$4s	   &      $^3$P$^o_{0}$    &	924.9250   &	 924.8705    &     924.8384   &    1.559-14   \\
   63   &   2s$^2$2p$^5$4s	   &      $^3$P$^o_{1}$    &	924.9937   &	 924.9393    &     924.9080   &    1.224-14   \\
   64   &   2s$^2$2p$^5$4p	   &      $^3$D$  _{1}$    &	929.2920   &	 929.2442    &     929.1628   &    1.282-14   \\
   65   &   2s$^2$2p$^5$4p	   &      $^3$P$  _{0}$    &	930.3910   &	 930.3735    &     930.2664   &    1.324-14   \\
   66   &   2s$^2$2p$^5$5s	   &      $^3$P$^o_{2}$    &	934.6085   &	 934.5911    &     934.4734   &    1.974-14   \\
   67   &   2s$^2$2p$^5$5s	   &      $^1$P$^o_{1}$    &	934.7035   &	 934.6879    &     934.5690   &    1.248-14   \\
   68   &   2s$^2$2p$^5$5p	   &      $^3$P$  _{1}$    &	936.8501   &	 936.8377    &     936.6935   &    1.634-14   \\
   69   &   2s$^2$2p$^5$5p	   &      $^3$D$  _{2}$    &	936.8745   &	 936.8627    &     936.7182   &    1.632-14   \\
   70   &   2s$^2$2p$^5$4p	   &      $^3$S$  _{1}$    &	940.1813   &	 940.1295    &     940.0616   &    2.020-14   \\
   71   &   2s$^2$2p$^5$4p	   &      $^1$D$  _{2}$    &	940.2343   &	 940.1822    &     940.1141   &    1.991-14   \\
   72   &   2s$^2$2p$^5$5p	   &      $^3$D$  _{3}$    &	942.2605   &	 942.2441    &     942.1071   &    2.413-14   \\
   73   &   2s$^2$2p$^5$5p	   &      $^1$P$  _{1}$    &	942.2805   &	 942.2644    &     942.1251   &    2.363-14   \\
   74   &   2s$^2$2p$^5$5p	   &      $^3$P$  _{2}$    &	942.4286   &	 942.4154    &     942.2751   &    2.426-14   \\
   75   &   2s$^2$2p$^5$5p	   &      $^1$S$  _{0}$    &	943.0864   &	 943.1085    &     942.9294   &    2.485-14   \\
   76   &   2s$^2$2p$^5$5d	   &      $^3$P$^o_{0}$    &	944.4309   &	 944.4140    &     944.2683   &    1.179-14   \\
   77   &   2s$^2$2p$^5$5d	   &      $^3$P$^o_{1}$    &	944.5447   &	 944.5283    &     944.3829   &    8.060-15   \\
   78   &   2s$^2$2p$^5$5d	   &      $^3$F$^o_{3}$    &	944.5542   &	 944.5392    &     944.3904   &    1.189-14   \\
   79   &   2s$^2$2p$^5$5d	   &      $^3$D$^o_{2}$    &	944.5899   &	 944.7027    &     944.4601   &    1.024-14   \\
   80   &   2s$^2$2p$^5$4d	   &      $^3$F$^o_{2}$    &	944.8201   &	 944.6487    &     944.6595   &    9.973-15   \\
   81   &   2s$^2$2p$^5$5d	   &      $^3$F$^o_{4}$    &	945.8519   &	 945.8335    &     945.6893   &    1.175-14   \\
   82   &   2s$^2$2p$^5$4d	   &      $^3$D$^o_{1}$    &	945.1744   &	 945.1335    &     945.0620   &    3.048-15   \\
   83   &   2s$^2$2p$^5$5d	   &      $^1$D$^o_{2}$    &	945.9508   &	 945.9281    &     945.7634   &    1.162-14   \\
   84   &   2s$^2$2p$^5$5d	   &      $^3$D$^o_{3}$    &	946.0654   &	 946.0513    &     945.8992   &    1.177-14   \\
   85   &   2s$^2$2p$^5$5d	   &      $^1$P$^o_{1}$    &	946.6542   &	 946.6491    &     946.4298   &    1.132-15   \\
   86   &   2s$^2$2p$^5$5f	   &      $^3$D$  _{1}$    &	947.1088   &	 947.0832    &     946.9442   &    7.718-15   \\
   87   &   2s$^2$2p$^5$5f	   &      $^3$G$  _{4}$    &	947.1324   &	 947.1071    &     946.9693   &    7.850-15   \\
   88   &   2s$^2$2p$^5$5f	   &      $^3$D$  _{2}$    &	947.1976   &	 947.1733    &     947.0339   &    7.735-15   \\
   89   &   2s$^2$2p$^5$5f	   &      $^3$F$  _{3}$    &	947.2405   &	 947.2171    &     947.0771   &    7.802-15   \\
   90   &   2s$^2$2p$^5$5f	   &      $^3$G$  _{5}$    &	947.7248   &	 947.6986    &     947.5602   &    8.058-15   \\
   91   &   2s$^2$2p$^5$5f	   &      $^1$D$  _{2}$    &	947.7994   &	 947.7756    &     947.3495   &    7.837-15   \\
   92   &   2s$^2$2p$^5$5f	   &      $^1$F$  _{3}$    &	947.8079   &	 947.7831    &     947.4067   &    7.983-15   \\
   93   &   2s$^2$2p$^5$5f	   &      $^3$F$  _{4}$    &	947.8519   &	 947.8279    &     947.6876   &    8.042-15   \\
   94   &   2s$^2$2p$^5$5g	   &      $^3$F$^o_{2}$    &	947.9245   &	 947.8952    &     947.6363   &    1.375-14   \\
   95   &   2s$^2$2p$^5$5g	   &      $^3$H$^o_{5}$    &	947.9293   &	 947.8998    &     947.7476   &    1.368-14   \\
   96   &   2s$^2$2p$^5$5g	   &      $^3$G$^o_{3}$    &	947.9762   &	 947.9468    &     947.6428   &    1.372-14   \\
   97   &   2s$^2$2p$^5$5g	   &      $^3$G$^o_{4}$    &	947.9948   &	 947.9654    &     947.8136   &    1.369-14   \\
   98   &   2s$^2$2p$^5$5g	   &      $^1$F$^o_{3}$    &	948.2936   &	 948.2646    &     947.7951   &    1.389-14   \\
   99   &   2s$^2$2p$^5$5g	   &      $^3$H$^o_{6}$    &	948.2901   &	 948.2606    &     948.1082   &    1.384-14   \\
  100   &   2s$^2$2p$^5$5g	   &      $^1$G$^o_{4}$    &	948.3391   &	 948.3096    &     948.1578   &    1.388-14   \\
  101   &   2s$^2$2p$^5$5g	   &      $^3$G$^o_{5}$    &	948.3543   &	 948.3248    &     948.1729   &    1.385-14   \\
  102   &   2s$^2$2p$^5$4d	   &      $^3$P$^o_{2}$    &	947.5010   &	 947.4496    &     947.7432   &    8.448-15   \\
  103   &   2s$^2$2p$^5$4d	   &      $^1$F$^o_{3}$    &	947.5560   &	 947.5041    &     948.1106   &    8.476-15   \\
  104   &   2s$^2$2p$^5$4f	   &      $^3$G$  _{3}$    &	950.0515   &	 949.9818    &     949.8944   &    4.098-15   \\
  105   &   2s$^2$2p$^5$4f	   &      $^3$F$  _{2}$    &	950.1189   &	 950.0504    &     949.9632   &    4.037-15   \\
  106   &   2s$^2$2p$^5$4f	   &      $^3$D$  _{3}$    &	951.2556   &	 951.1855    &     951.0958   &    4.182-15   \\
  107   &   2s$^2$2p$^5$4f	   &      $^1$G$  _{4}$    &	951.2479   &	 951.1779    &     951.0889   &    4.213-15   \\
  108   &   2s2p$^6$4s  	   &      $^3$S$  _{1}$    &	959.9474   &	 960.1425    &     959.6874   &    1.194-14   \\
  109   &   2s2p$^6$4s  	   &      $^1$S$  _{0}$    &	960.7144   &	 960.9429    &     960.4493   &    1.204-14   \\
  110   &   2s2p$^6$4p  	   &      $^3$P$^o_{0}$    &	964.4513   &	 964.6587    &     964.1518   &    1.035-14   \\
  111   &   2s2p$^6$4p  	   &      $^3$P$^o_{1}$    &	964.4817   &	 964.6892    &     964.1823   &    4.378-15   \\
  112   &   2s2p$^6$4p  	   &      $^3$P$^o_{2}$    &	975.2367   &	 975.4349    &     974.9460   &    1.444-14   \\
  113   &   2s2p$^6$4p  	   &      $^1$P$^o_{1}$    &	975.3897   &	 975.5897    &     975.0979   &    3.471-15   \\
  114   &   2s2p$^6$4d  	   &      $^3$D$  _{1}$    &	979.8084   &	 980.0128    &     979.4810   &    7.430-15   \\
  115   &   2s2p$^6$4d  	   &      $^3$D$  _{2}$    &	979.9419   &	 980.1481    &     979.6167   &    7.388-15   \\
  116   &   2s2p$^6$4d  	   &      $^3$D$  _{3}$    &	982.4031   &	 982.6002    &     982.0808   &    7.193-15   \\
  117   &   2s2p$^6$4d  	   &      $^1$D$  _{2}$    &	982.7892   &	 982.9937    &     982.4622   &    7.118-15   \\
  118   &   2s2p$^6$4f  	   &      $^3$F$^o_{2}$    &	984.9482   &	 985.1127    &     984.6082   &    3.760-15   \\
  119   &   2s2p$^6$4f  	   &      $^3$F$^o_{3}$    &	984.9306   &	 985.0948    &     984.5975   &    3.767-15   \\
  120   &   2s2p$^6$4f  	   &      $^3$F$^o_{4}$    &	986.0831   &	 986.2468    &     985.7484   &    3.853-15   \\
  121   &   2s2p$^6$4f  	   &      $^1$F$^o_{3}$    &	986.1675   &	 986.3313    &     985.8256   &    3.868-15   \\
\hline  											      
			      							   					       

\\
\end{longtable}

\clearpage
\newpage

\begin{longtable}{@{\extracolsep\fill}rllrrrr@{}}
\caption{Energies (Ryd) for the lowest  117 levels of Re~LXVI and their lifetimes ($\tau$, s).   $a{\pm}b \equiv a{\times}$10$^{{\pm}b}$. See page\ \pageref{tbl3te} for Explanation of Tables} 
Index  &     Configuration                        & Level     & GRASP$^a$   & GRASP$^b$   &   FAC   &  $\tau$ (s)  \\  \\
\hline\\
\endfirsthead\\
\caption[]{(continued)}
Index  &     Configuration                        & Level     & GRASP$^a$   & GRASP$^b$   &   FAC   &  $\tau$ (s)   \\  \\
\hline\\
\endhead
    1  &  2s$^2$2p$^6$  	 &    $^1$S$  _{0}$   &     0.0000   &      0.0000   &      0.0000  &  ........   \\
    2  &  2s$^2$2p$^5$3s	 &    $^3$P$^o_{2}$   &   626.9177   &    626.8857   &    627.0385  &  2.060-10   \\
    3  &  2s$^2$2p$^5$3s	 &    $^1$P$^o_{1}$   &   627.5527   &    627.5184   &    627.6896  &  6.115-15   \\
    4  &  2s$^2$2p$^5$3p	 &    $^3$P$  _{1}$   &   638.2365   &    638.2366   &    638.2357  &  1.410-11   \\
    5  &  2s$^2$2p$^5$3p	 &    $^3$D$  _{2}$   &   638.3597   &    638.3612   &    638.3640  &  9.599-13   \\
    6  &  2s$^2$2p$^5$3p	 &    $^1$P$  _{1}$   &   667.9399   &    667.9257   &    667.9511  &  5.542-13   \\
    7  &  2s$^2$2p$^5$3p	 &    $^3$D$  _{3}$   &   667.9053   &    667.8903   &    667.9180  &  5.299-13   \\
    8  &  2s$^2$2p$^5$3p	 &    $^3$P$  _{2}$   &   668.8821   &    668.8759   &    668.9069  &  3.343-13   \\
    9  &  2s$^2$2p$^5$3p	 &    $^1$S$  _{0}$   &   673.2037   &    673.2555   &    673.2647  &  3.668-13   \\
   10  &  2s$^2$2p$^5$3d	 &    $^3$P$^o_{0}$   &   679.8437   &    679.8178   &    679.7631  &  7.246-13   \\
   11  &  2s$^2$2p$^5$3d	 &    $^3$P$^o_{1}$   &   680.5899   &    680.5647   &    680.5131  &  1.074-14   \\
   12  &  2s$^2$2p$^5$3d	 &    $^3$F$^o_{3}$   &   680.6398   &    680.6150   &    680.5673  &  6.908-13   \\
   13  &  2s$^2$2p$^5$3d	 &    $^3$D$^o_{2}$   &   681.2539   &    681.2308   &    681.1772  &  6.589-13   \\
   14  &  2s$^2$2p$^5$3d	 &    $^3$F$^o_{4}$   &   687.1579   &    687.1255   &    687.0831  &  5.961-12   \\
   15  &  2s$^2$2p$^5$3d	 &    $^1$D$^o_{2}$   &   687.6964   &    687.6663   &    687.6205  &  3.217-12   \\
   16  &  2s$^2$2p$^5$3d	 &    $^3$D$^o_{3}$   &   688.4477   &    688.4206   &    688.3743  &  5.452-12   \\
   17  &  2s$^2$2p$^5$3d	 &    $^1$P$^o_{1}$   &   690.6351   &    690.6284   &    690.5483  &  3.400-16   \\
   18  &  2s$^2$2p$^5$3s	 &    $^3$P$^o_{0}$   &   735.3052   &    735.2385   &    735.4873  &  2.624-11   \\
   19  &  2s$^2$2p$^5$3s	 &    $^3$P$^o_{1}$   &   735.5551   &    735.4860   &    735.7439  &  2.409-14   \\
   20  &  2s$^2$2p$^5$3p	 &    $^3$D$  _{1}$   &   746.3184   &    746.2737   &    746.3759  &  1.374-11   \\
   21  &  2s$^2$2p$^5$3p	 &    $^3$P$  _{0}$   &   749.9908   &    749.9872   &    750.0742  &  2.916-12   \\
   22  &  2s2p$^6$3s		 &    $^3$S$  _{1}$   &   771.5774   &    771.8000   &    771.4939  &  4.344-14   \\
   23  &  2s2p$^6$3s		 &    $^1$S$  _{0}$   &   774.6986   &    775.0469   &    774.6003  &  4.075-14   \\
   24  &  2s$^2$2p$^5$3p	 &    $^1$D$  _{2}$   &   776.6224   &    776.5715   &    776.7009  &  3.186-13   \\
   25  &  2s$^2$2p$^5$3p	 &    $^3$S$  _{1}$   &   777.1101   &    777.0940   &    777.1450  &  1.673-13   \\
   26  &  2s2p$^6$3p		 &    $^3$P$^o_{0}$   &   783.6231   &    783.8923   &    783.3903  &  3.846-14   \\
   27  &  2s2p$^6$3p		 &    $^3$P$^o_{1}$   &   783.4883   &    783.7456   &    783.2754  &  1.318-15   \\
   28  &  2s$^2$2p$^5$3d	 &    $^3$F$^o_{2}$   &   789.0007   &    788.9369   &    788.9925  &  6.591-13   \\
   29  &  2s$^2$2p$^5$3d	 &    $^3$D$^o_{1}$   &   790.9567   &    790.9173   &    790.9205  &  6.601-16   \\
   30  &  2s$^2$2p$^5$3d	 &    $^3$P$^o_{2}$   &   796.0266   &    795.9605   &    796.0120  &  3.033-12   \\
   31  &  2s$^2$2p$^5$3d	 &    $^1$F$^o_{3}$   &   796.2795   &    796.2127   &    796.2659  &  4.794-12   \\
   32  &  2s2p$^6$3p		 &    $^3$P$^o_{2}$   &   813.4503   &    813.6977   &    813.2350  &  3.573-14   \\
   33  &  2s2p$^6$3p		 &    $^1$P$^o_{1}$   &   813.9029   &    814.1484   &    813.6944  &  2.030-15   \\
   34  &  2s2p$^6$3d		 &    $^3$D$  _{1}$   &   825.8034   &    826.0893   &    825.4731  &  3.989-14   \\
   35  &  2s2p$^6$3d		 &    $^3$D$  _{2}$   &   826.1964   &    826.4821   &    825.8769  &  3.102-14   \\
   36  &  2s2p$^6$3d		 &    $^3$D$  _{3}$   &   832.5962   &    832.8699   &    832.2704  &  3.953-14   \\
   37  &  2s2p$^6$3d		 &    $^1$D$  _{2}$   &   833.7414   &    834.0198   &    833.4326  &  2.684-14   \\
   38  &  2s$^2$2p$^5$4s	 &    $^3$P$^o_{2}$   &   879.3542   &    879.3475   &    879.2729  &  1.372-14   \\
   39  &  2s$^2$2p$^5$4s	 &    $^1$P$^o_{1}$   &   879.5648   &    879.5595   &    879.4860  &  7.709-15   \\
   40  &  2s$^2$2p$^5$4p	 &    $^3$P$  _{1}$   &   884.0291   &    884.0304   &    883.8966  &  1.133-14   \\
   41  &  2s$^2$2p$^5$4p	 &    $^3$D$  _{2}$   &   884.0644   &    884.0660   &    883.9327  &  1.125-14   \\
   42  &  2s$^2$2p$^5$4p	 &    $^3$D$  _{3}$   &   896.2168   &    896.2112   &    896.0924  &  1.813-14   \\
   43  &  2s$^2$2p$^5$4p	 &    $^1$P$  _{1}$   &   896.2418   &    896.2364   &    896.1161  &  1.803-14   \\
   44  &  2s$^2$2p$^5$4p	 &    $^3$P$  _{2}$   &   896.5780   &    896.5753   &    896.4548  &  1.820-14   \\
   45  &  2s$^2$2p$^5$4p	 &    $^1$S$  _{0}$   &   898.0030   &    898.0605   &    897.8787  &  1.878-14   \\
   46  &  2s$^2$2p$^5$4d	 &    $^3$P$^o_{0}$   &   900.7853   &    900.7814   &    900.6281  &  7.746-15   \\
   47  &  2s$^2$2p$^5$4d	 &    $^3$P$^o_{1}$   &   901.0528   &    901.0594   &    900.8948  &  6.068-15   \\
   48  &  2s$^2$2p$^5$4d	 &    $^3$F$^o_{3}$   &   901.0568   &    901.0548   &    900.8978  &  7.774-15   \\
   49  &  2s$^2$2p$^5$4d	 &    $^3$D$^o_{2}$   &   901.2789   &    901.2791   &    901.1180  &  7.809-15   \\
   50  &  2s$^2$2p$^5$4d	 &    $^3$F$^o_{4}$   &   903.8995   &    903.8924   &    903.7426  &  7.464-15   \\
   51  &  2s$^2$2p$^5$4d	 &    $^1$D$^o_{2}$   &   904.0977   &    904.0928   &    903.9384  &  7.462-15   \\
   52  &  2s$^2$2p$^5$4d	 &    $^3$D$^o_{3}$   &   904.3662   &    904.3640   &    904.2053  &  7.508-15   \\
   53  &  2s$^2$2p$^5$4d	 &    $^1$P$^o_{1}$   &   905.1290   &    905.1370   &    904.9529  &  8.436-16   \\
   54  &  2s$^2$2p$^5$4f	 &    $^3$D$  _{1}$   &   906.5567   &    906.5339   &    906.3891  &  3.534-15   \\
   55  &  2s$^2$2p$^5$4f	 &    $^3$G$  _{4}$   &   906.5720   &    906.5491   &    906.4070  &  3.624-15   \\
   56  &  2s$^2$2p$^5$4f	 &    $^3$F$  _{2}$   &   906.7342   &    906.7121   &    906.5685  &  3.572-15   \\
   57  &  2s$^2$2p$^5$4f	 &    $^3$F$  _{3}$   &   906.8125   &    906.7905   &    906.6475  &  3.619-15   \\
   58  &  2s$^2$2p$^5$4f	 &    $^3$G$  _{5}$   &   907.8870   &    907.8632   &    907.7195  &  3.720-15   \\
   59  &  2s$^2$2p$^5$4f	 &    $^1$D$  _{2}$   &   908.0164   &    907.9953   &    907.8514  &  3.613-15   \\
   60  &  2s$^2$2p$^5$4f	 &    $^1$F$  _{3}$   &   908.0804   &    908.0575   &    907.9121  &  3.689-15   \\
   61  &  2s$^2$2p$^5$4f	 &    $^3$F$  _{4}$   &   908.1513   &    908.1287   &    907.9849  &  3.730-15   \\
   62  &  2s$^2$2p$^5$4s	 &    $^3$P$^o_{0}$   &   987.8809   &    987.8259   &    987.8137  &  1.358-14   \\
   63  &  2s$^2$2p$^5$4s	 &    $^3$P$^o_{1}$   &   987.9479   &    987.8930   &    987.8815  &  1.071-14   \\
   64  &  2s$^2$2p$^5$5s	 &    $^3$P$^o_{2}$   &   991.8946   &    991.8800   &    991.7648  &  1.720-14   \\
   65  &  2s$^2$2p$^5$5s	 &    $^1$P$^o_{1}$   &   991.9954   &    991.9826   &    991.8661  &  1.101-14   \\
   66  &  2s$^2$2p$^5$4p	 &    $^3$D$  _{1}$   &   992.4333   &    992.3851   &    992.3132  &  1.118-14   \\
   67  &  2s$^2$2p$^5$5p	 &    $^3$P$  _{1}$   &   994.2408   &    994.2316   &    994.0842  &  1.428-14   \\
   68  &  2s$^2$2p$^5$5p	 &    $^3$D$  _{2}$   &   994.2592   &    994.2504   &    994.1030  &  1.425-14   \\
   69  &  2s$^2$2p$^5$4p	 &    $^3$P$  _{0}$   &   993.5548   &    993.5349   &    993.4396  &  1.157-14   \\
   70  &  2s$^2$2p$^5$5p	 &    $^1$P$  _{1}$   &  1000.3856   &   1000.3720   &   1000.2325  &  2.125-14   \\
   71  &  2s$^2$2p$^5$5p	 &    $^3$D$  _{3}$   &  1000.3821   &   1000.3680   &   1000.2295  &  2.165-14   \\
   72  &  2s$^2$2p$^5$5p	 &    $^3$P$  _{2}$   &  1000.5573   &   1000.5470   &   1000.4046  &  2.175-14   \\
   73  &  2s$^2$2p$^5$5p	 &    $^1$S$  _{0}$   &  1001.2710   &   1000.2990   &   1001.1145  &  2.215-14   \\
   74  &  2s$^2$2p$^5$5d	 &    $^3$P$^o_{0}$   &  1002.6284   &   1002.6140   &   1002.4647  &  1.046-14   \\
   75  &  2s$^2$2p$^5$5d	 &    $^3$P$^o_{1}$   &  1002.7549   &   1002.7420   &   1002.5904  &  8.624-15   \\
   76  &  2s$^2$2p$^5$5d	 &    $^3$F$^o_{3}$   &  1002.7543   &   1002.7420   &   1002.5894  &  1.055-14   \\
   77  &  2s$^2$2p$^5$5d	 &    $^3$D$^o_{2}$   &  1002.8596   &   1002.8490   &   1002.6931  &  1.058-14   \\
   78  &  2s$^2$2p$^5$5d	 &    $^3$F$^o_{4}$   &  1004.2258   &   1004.2100   &   1004.0622  &  1.039-14   \\
   79  &  2s$^2$2p$^5$5d	 &    $^1$D$^o_{2}$   &  1004.3157   &   1004.3010   &   1004.1505  &  1.035-14   \\
   80  &  2s$^2$2p$^5$5d	 &    $^3$D$^o_{3}$   &  1004.4484   &   1004.4370   &   1004.2812  &  1.041-14   \\
   81  &  2s$^2$2p$^5$5d	 &    $^1$P$^o_{1}$   &  1004.7574   &   1004.7530   &   1004.5826  &  2.070-15   \\
   82  &  2s$^2$2p$^5$5f	 &    $^3$D$  _{1}$   &  1005.5257   &   1005.5030   &   1004.7058  &  6.824-15   \\
   83  &  2s$^2$2p$^5$5f	 &    $^3$G$  _{4}$   &  1005.5440   &   1005.5210   &   1005.3800  &  6.936-15   \\
   84  &  2s$^2$2p$^5$5f	 &    $^3$D$  _{2}$   &  1005.6200   &   1005.5990   &   1004.7391  &  6.846-15   \\
   85  &  2s$^2$2p$^5$5f	 &    $^3$F$  _{3}$   &  1005.6583   &   1005.6370   &   1005.4938  &  6.897-15   \\
   86  &  2s$^2$2p$^5$4p	 &    $^3$S$  _{1}$   &  1004.8153   &   1004.7630   &   1005.4534  &  1.810-14   \\
   87  &  2s$^2$2p$^5$4p	 &    $^1$D$  _{2}$   &  1004.8405   &   1004.7880   &   1005.3591  &  1.774-14   \\
   88  &  2s$^2$2p$^5$5f	 &    $^3$G$  _{5}$   &  1006.2165   &   1006.1930   &   1006.0510  &  7.133-15   \\
   89  &  2s$^2$2p$^5$5f	 &    $^1$D$  _{2}$   &  1006.3057   &   1006.2840   &   1006.1353  &  6.932-15   \\
   90  &  2s$^2$2p$^5$5f	 &    $^1$F$  _{3}$   &  1006.3060   &   1006.2840   &   1006.1398  &  7.074-15   \\
   91  &  2s$^2$2p$^5$5f	 &    $^3$F$  _{4}$   &  1006.3493   &   1006.3280   &   1006.1841  &  7.120-15   \\
   92  &  2s$^2$2p$^5$5g	 &    $^3$F$^o_{2}$   &  1006.4237   &   1006.3970   &   1006.2950  &  1.213-14   \\
   93  &  2s$^2$2p$^5$5g	 &    $^3$H$^o_{5}$   &  1006.4268   &   1006.4000   &   1006.2441  &  1.208-14   \\
   94  &  2s$^2$2p$^5$5g	 &    $^3$G$^o_{3}$   &  1006.4772   &   1006.4500   &   1006.2414  &  1.211-14   \\
   95  &  2s$^2$2p$^5$5g	 &    $^3$G$^o_{4}$   &  1006.4954   &   1006.4680   &   1006.3131  &  1.209-14   \\
   96  &  2s$^2$2p$^5$5g	 &    $^1$F$^o_{3}$   &  1006.8358   &   1006.8090   &   1006.6533  &  1.229-14   \\
   97  &  2s$^2$2p$^5$5g	 &    $^3$H$^o_{6}$   &  1006.8353   &   1006.8080   &   1006.6525  &  1.223-14   \\
   98  &  2s$^2$2p$^5$5g	 &    $^1$G$^o_{4}$   &  1006.8874   &   1006.8600   &   1006.7051  &  1.226-14   \\
   99  &  2s$^2$2p$^5$5g	 &    $^3$G$^o_{5}$   &  1006.9022   &   1006.8750   &   1006.7198  &  1.224-14   \\
  100  &  2s$^2$2p$^5$4d	 &    $^3$F$^o_{2}$   &  1009.5648   &   1009.5140   &   1009.4219  &  7.782-15   \\
  101  &  2s$^2$2p$^5$4d	 &    $^3$D$^o_{1}$   &  1010.2461   &   1010.2050   &   1010.0911  &  1.102-15   \\
  102  &  2s$^2$2p$^5$4d	 &    $^3$P$^o_{2}$   &  1012.6120   &   1012.5590   &   1012.4694  &  7.453-15   \\
  103  &  2s$^2$2p$^5$4d	 &    $^1$F$^o_{3}$   &  1012.6882   &   1012.6350   &   1012.5446  &  7.486-15   \\
  104  &  2s$^2$2p$^5$4f	 &    $^3$G$  _{3}$   &  1015.2558   &   1015.1850   &   1015.1055  &  3.620-15   \\
  105  &  2s$^2$2p$^5$4f	 &    $^3$D$  _{2}$   &  1015.3243   &   1015.2550   &   1015.1756  &  3.564-15   \\
  106  &  2s$^2$2p$^5$4f	 &    $^3$D$  _{3}$   &  1016.6165   &   1016.5450   &   1016.4634  &  3.700-15   \\
  107  &  2s$^2$2p$^5$4f	 &    $^1$G$  _{4}$   &  1016.6095   &   1016.5380   &   1016.4572  &  3.729-15   \\
  108  &  2s2p$^6$4s		 &    $^3$S$  _{1}$   &  1024.4476   &   1024.6480   &   1024.1239  &  1.008-14   \\
  109  &  2s2p$^6$4s		 &    $^1$S$  _{0}$   &  1025.2527   &   1025.4870   &   1024.9232  &  1.020-14   \\
  110  &  2s2p$^6$4p		 &    $^3$P$^o_{0}$   &  1029.1530   &   1029.3660   &   1028.7793  &  8.790-15   \\
  111  &  2s2p$^6$4p		 &    $^3$P$^o_{1}$   &  1029.1804   &   1029.3930   &   1028.8066  &  3.788-15   \\
  112  &  2s2p$^6$4p		 &    $^3$P$^o_{2}$   &  1041.4055   &   1041.6080   &   1041.0422  &  1.240-14   \\
  113  &  2s2p$^6$4p		 &    $^1$P$^o_{1}$   &  1041.5666   &   1041.7710   &   1041.2020  &  3.071-15   \\
  114  &  2s2p$^6$4d		 &    $^3$D$  _{1}$   &  1046.1365   &   1046.3460   &   1045.7299  &  6.471-15   \\
  115  &  2s2p$^6$4d		 &    $^3$D$  _{2}$   &  1046.2701   &   1046.4810   &   1045.8673  &  6.420-15   \\
  116  &  2s2p$^6$4d		 &    $^3$D$  _{3}$   &  1049.0724   &   1049.2740   &   1048.6721  &  6.245-15   \\
  117  &  2s2p$^6$4d		 &    $^1$D$  _{2}$   &  1049.4720   &   1049.6810   &   1049.0669  &  6.174-15   \\
\hline  											      
			      							   					       

\\
\end{longtable}

\clearpage
\newpage

\end{document}